\documentclass[iop]{emulateapj}
\usepackage{graphicx}
\usepackage{epstopdf}
\usepackage{amsmath, amsthm, amssymb}
\usepackage{multirow}
\usepackage{bbding}

\newcommand{\Starlab}{\textsc{Starlab} }
\newcommand{\AMUSE}{\textsc{AMUSE} }
\newcommand{\AMUSEper}{\textsc{AMUSE}. }
\newcommand{\na}{NewA}

\newcommand{\rr}[1]{{#1}}
\newcommand{\rrtwo}[1]{{#1}}

\begin{document}

\title{Simulating Star Clusters with the AMUSE Software Framework:
  I. Dependence of Cluster Lifetimes on Model Assumptions and Cluster
  Dissolution Modes}

\author{Alfred J. Whitehead}
\email{alf.whitehead@drexel.edu}
\author{Stephen L. W. McMillan}
\affiliation{Drexel University, Philadelphia, PA, 19104, USA}

\author{Enrico Vesperini}
\affiliation{Indiana University, Bloomington, IN, 47405, USA}

\author{Simon \surname{Portegies Zwart}}
\affiliation{Leiden Observatory, Leiden University, Leiden, The Netherlands}

\shorttitle{Simulating Star Clusters with AMUSE}
\shortauthors{Whitehead et al.}

\pacs{98.20.-d, 98.10.+z, 95.75.Pq}

\begin{abstract}
We perform a series of simulations of evolving star clusters using
\AMUSE (the Astrophysical Multipurpose Software Environment), a new
community-based multi-physics simulation package, and compare our
results to existing work.  These simulations model a star cluster beginning
with a King model distribution and a selection of power-law initial mass functions,
and contain a tidal cut-off.  They are evolved using collisional stellar dynamics
and include mass loss due to stellar evolution.  After \rrtwo{studying and understanding} that the 
differences between \AMUSE results and \rr{results from previous studies} are understood, we 
explored the variation in cluster lifetimes due to the random realization noise
introduced by transforming a King model to specific initial conditions.  This random 
realization noise can affect the lifetime of a simulated star cluster by up to 30\%.
Two modes of star cluster dissolution were identified: a mass
evolution curve that contains a run-away cluster dissolution with a sudden loss of mass, and a dissolution mode
that does not contain this feature.  We refer to these dissolution modes as ``dynamical'' and ``relaxation'' dominated respectively.  
For Salpeter-like initial mass functions, we determined
the boundary between these two modes in terms of the dynamical and relaxation timescales.
\end{abstract}

\maketitle

\section{Introduction}

Star clusters are natural laboratories for many astrophysical
processes.  In the simplest description, cluster stars may be thought
of as being (almost) coeval point masses---an N-body system---and
their motion traces their mutual gravitation and the possible
influence of an external galactic tidal field.  In more complex
situations, stars evolve, gas may accrete into the cluster, new stars
may form out of that gas, and the gas may be expelled from the cluster
quickly by supernovae or more slowly by radiation pressure and stellar
winds.  A typical cluster is subject to several long-term mass-loss
processes, including losses due to stellar evolution and 
\rr{removal} of the outermost stars \rr{by the galaxy's tidal field}.  These 
processes compete with
relaxation processes to define the equilibrium state of the cluster.

Setting aside the complexities of intracluster gas, simple models
combining a few basic physical processes---stellar dynamics, stellar
evolution, and tidal effects---have proved very useful in the study of
star clusters.  These simulations combine differing treatments of
multiple physical processes, and must be carefully calibrated to
ensure their reliability.  \citet{CW90} (noted as ``CW'' throughout
this paper) combined a simple stellar evolution prescription with
Fokker-Planck simulations of stellar dynamics and a highly idealized
tidal field to produce a seminal ``baseline'' set of cluster
simulations, starting from King \citep{K66} initial models.  This
survey, together with subsequent studies by \citet{F95}, \citet{A98},
and \citet{TPZ00} (abbreviated as ``TPZ'' for the remainder of this
paper), using other formulations of stellar evolution and both N-body
and Fokker--Planck treatments of stellar dynamics, have resulted in
comparative catalogs of parameter space that now serve as tests of any
new code.

Part of the purpose of this paper is to validate parts of
{\AMUSE}\footnote{http://amusecode.org}---the Astrophysical Multipurpose 
Simulation Environment---against known results and then to show new applications
of the framework to stellar cluster dynamics.
{\AMUSE} is a new software framework designed for simulations of dense stellar
systems, inspired by the earlier \textsc{MUSE} project described by
\citet{H08} and \citet{PZ09}.  A detailed technical account of
{\AMUSE} is beyond the scope of this article \citep[see][]{M11a,PZ13}. A summary is
presented in \S\ref{Sec:Computation} to provide the reader with some
context on the software used.

We set out to test {\AMUSE} against known results, but found that
comparing different simulations at any meaningful level of precision
is a non-trivial task.  \rrtwo{In order to accomplish this
goal}, we employ an N-body stellar
dynamics code, several stellar evolution codes, and a simple \rr{escaper removal} 
algorithm as the three basic simulation components, and
compare {\AMUSE} with the results of TPZ. 

This line of inquiry led to a description of the dissolution modes of King models
within a tidal \rrtwo{cut-off}.  We demonstrate that competition between the relaxation,
dynamical and stellar evolution timescales leads to a split
between dissolutions dominated by relaxation processes and those dominated
by dynamical processes.  By sampling the relevant timescales, the boundary is 
mapped.

We also generate a
comparison of different stellar evolution codes linked to the same
dynamics code and run against the same initial conditions,
demonstrating that the specifics of the choice of stellar evolution
recipes are amplified by the stellar dynamics, and impact the results
of the simulation.

The structure of this paper is as
follows: in \S\ref{Sec:Computation} we describe \AMUSE and its
specific use for the dissolving star cluster problem.  This is followed by
\S\ref{physmodel} where the physical model, including details of the CW
stellar evolution approximation, are detailed.  \S\ref{resultssec} contains the
validation of \AMUSE runs (\S\ref{validationsec} and \S\ref{sec:comparisontpz}), 
a study of the consequences of the variance 
in initial conditions on simulations (\S\ref{sec:sources}), an exploration of the
types of dissolution which can disrupt a King model (\S\ref{sec:typesofdis}), 
and a direct comparison of stellar evolution
codes (\S\ref{secomp}).  Finally, \S\ref{futurework} summarizes the results and proposes future
work.

\rr{This paper is the first in a series of papers describing work with
  \AMUSEper In this series we will lay the groundwork for future
  studies by demonstrating that \AMUSE can reproduce well-known
  published results.  Future work will explore various types of N-body
  codes (direct integration, tree, etc.) as well as the inclusion of
  binaries and multiple stars.

The series begins with a relatively simple model (single stars in a cluster
using a tidal \rrtwo{cut-off}), and is intended to progress to more realistic models
in later work.  It is important to establish the reliability of the \AMUSE framework
through comparison with existing work.  \rrtwo{We} can demonstrate the utility
of the modular framework along the way by conducting comparisons between
codes that were prohibitively difficult without out \rrtwo{it}.}

\section{Computational Framework}\label{Sec:Computation}
Historically, astrophysical simulation codes have been constructed by a
single author or by a small group working closely together.  The
typical course of the development begins with a simple solver for a
specific physical problem (for example, an N-body integrator for a
collisionless system) and then gradually extends to cover more varied
physics (to continue the example, collisional physics or stellar
evolution effects might be added).  
\rr{This approach has been very successful, but is limited when it comes
time to compare codes and implementations, or to extend a simulation to include a new piece 
of physics (to continue the example again, radiative transfer processes may
need to be included).  In the case of comparison, the types of physics studied are
tightly coupled to a specific implementation.  It is non-trivial to change from
one stellar evolution recipe (to give one example) to another, unless the authors
of the code have included both recipes.  In the case of extension, the team of authors
behind the code may need to grow to bring in experts in the newly-required fields
of physics.}

\rr{Despite these difficulties, a number of very successful 
codes have been developed.}  Among these are the \textsc{Nbody} series
of codes \citep[for a review, see][]{A99}, \textsc{Gadget}
\citep{SYW01}, \textsc{Flash} \citep{F00} and
{\Starlab} \citep{PZ01}. \rr{Nevertheless, it is becoming clear
that the limits of the traditional 
approach are being reached.}  In
order for new physics to be added to these packages, the programmer
(or team of programmers) must be an expert in the new physics being
added, as well as in every physical domain already present in the
tightly coupled code.  \rr{This, combined with the difficulty of modifying
any existing physics in these packages, limits the
effectiveness of further work.} 

The {\AMUSE} philosophy is to move away from a general-purpose multiphysics ``solver'' and toward a suite of standardized special-purpose ``evolver'' modules.  Each evolver knows about only a single 
physical domain, and is responsible for advancing a known system state through time by implementing the physics specific to that domain.  In particular, an evolver is not expected to take into account any 
physics outside its own domain in its calculations.

The {\AMUSE} standard defines four physical domains of interest:
gravitational dynamics, stellar evolution, hydrodynamics and radiative
transfer.  A standard interface to an evolver is defined for each of
these domains.  For example, the stellar dynamics interface specifies
how particles are communicated to the evolver (added, removed,
updated) and how to make the evolver step forward a given number of
time units.  Similarly, the stellar evolution interface specifies how
to communicate star properties (mass, age, metallicity, etc.) to the
evolver and how to make the evolver advance to a specified
time.

All evolvers for a given physical domain are accessible within the {\AMUSE} environment through this standard interface.  This means that evolvers within a domain are interchangeable.  As shown in 
\S\ref{resultssec}, it is possible for a researcher to switch between several stellar evolution models to test the effect of changing the physical approximations used on the behaviour of the entire 
system.  The same is true of the other physical domains.  \rr{This decoupling of the 
underlying science codes from the simulation logic is powerful.  Users who
are not experts in the details of the scientific modules can ``mix and match''
reliable existing work to produce new types of simulations.}

Wherever possible, the {\AMUSE} approach is to re-use existing codes
instead of writing new ones.  This means that many special-purpose
stand-alone solvers can be turned into {\AMUSE} modules.  The
framework provides a quick and easy method for wrapping an existing
code in one of the standard interfaces and making it available within
the {\AMUSE} environment.  \rr{The decoupling of science codes is
a benefit to code authors, as they now need produce only a solver for
an individual physics domain in order to run a realistic simulation.  The
other physical domains, in which they may not be experts, can be ``borrowed''
from the \AMUSE community codes directly.}  At the discretion of the author, such a
module may also become part of the {\AMUSE} package distributed on the
web to interested researchers.  Alternately, it is possible to create
a ``private'' {\AMUSE} module which exists only on the author's
computers.

In order to make use of {\AMUSE}, the researcher writes a
``top-level'' script (using the Python scripting language) which
instantiates a set of evolvers relevant to the problem being studied.
All communication and \rrtwo{synchronization} between the evolvers are handled
by this script.  In this work, the top-level script creates a stellar
dynamics evolver (in our case, an N-body code) and a stellar evolution
evolver.  It then begins a loop in which dynamics and evolution are
advanced in tandem, with synchronization between them as needed.  It
also implements a tidal cut-off by removing escapers from the
simulation at fixed time intervals.

{\AMUSE} uses the message-passing interface \textsc{MPI} \citep[see,
for example,][]{W94} to allow each evolver to exist in its own
process, possibly in parallel and on a different machine than the
controlling Python script.  Each evolver is written in the
language of choice of its original author.  Already present in {\AMUSE}
are modules written in C, C++, Python, Fortran and Java.  \textsc{MPI} was
chosen based on the experience of \textsc{MUSE} (which used
\textsc{Swig} and \textsc{f2py} instead), and allows for both parallelization and for
each module to reside in its own, unique namespace.  \rr{\AMUSE is compatible with OpenMPI
or with MPICH2, or variants thereof.  In this work we have used the 
\textsc{MVAPICH2}\footnote{http://mvapich.cse.ohio-state.edu/overview/mvapich2/}
implementation of \textsc{MPI} because it supports the Infinibad networking
present on our GPU computing cluster.}  {\AMUSE} is also capable
of running on a grid for massively parallel calculations \citep{D12}.

\begin{table*}[htpb]
\begin{center}
\caption{{\AMUSE} Modules Used In This Work}
\begin{ruledtabular}
\begin{tabular}{ccc}
\startdata
Module & Type & Reference\\
\hline
\hline
\textsc{ph4} & N-body Dynamics & \citet{M11a} \\ \hline
\textsc{Sapporo} & GRAPE Emulator on GPU & \citet{G09} \\ \hline
\textsc{SSE} & Stellar Evolution & \citet{H00} \\ \hline
\textsc{EFT89} & Stellar Evolution & \citet{EFT89,EFT90} \\ \hline
\textsc{VSSE} & Stellar Evolution & \citet{CW90} \\ \hline
\textsc{SeBa} & Stellar Evolution & \citet{PZV96} \\ 
\end{tabular}
\end{ruledtabular}
\label{moduletable}
\end{center}
\end{table*}

Table \ref{moduletable} lists the specific {\AMUSE} modules used in this work.
The \textsc{ph4} evolver provides N-body dynamics using
\textsc{Sapporo}, a GRAPE emulator, for GPU
acceleration. \textsc{ph4} is an MPI-parallel, adaptive block
time-step, \textsc{GRAPE}-accelerated, $4^\mathrm{th}$-order Hermite
integrator, similar in construction to \textsc{phiGRAPE} \citep{H07}.
\textsc{SSE} provides stellar evolution. \textsc{knnCUDA}
\citep{G08a, G08b} is used to compute densities (using $12^{th}$ nearest
\rrtwo{neighbors}) in a \rrtwo{stand-alone} code similar to that described by
\citet{CH85}, but separate from {\AMUSE}.  This code uses an exact
algorithm to find all nearest \rrtwo{neighbors}, regardless of distance.

\rr{All modules used in this work were compiled for a 64-bit architecture
and use double precision floating point variables.  \AMUSE provides
unit conversion features to link the N-body code (using dimensionless
units) to the stellar evolution code (using physical units).  This linkage
occurs only in the time and mass variables, and a brief analysis shows
that no significant numerical error is expected.}

\rr{While we have used \textsc{Sapporo}, which in turn uses \textsc{CUDA}, to
emulate a \textsc{GRAPE} hardware accelerator, it is important to note that
this is not the only possible choice.  An actual \textsc{GRAPE} could be used,
as could the \textsc{sapporo\_light} module included in \AMUSE, which can run
with or without a GPU present.  Future versions of 
\textsc{Sapporo} will use OpenCL, which will allow use of the
\AMUSE packages on AMD-based GPUs and other devices supporting that
open standard.}

Additionally, the \textsc{EFT89} module, the \textsc{SeBa} module and the
\textsc{VSSE} (Very Simple Stellar Evolution) module were used to
provide the simplified stellar evolution models explored in section
\ref{resultssec}.  \textsc{VSSE} was written specifically for this
work, and is designed to allow a researcher to easily add simple
analytical stellar evolution models, here the prescription of CW (see
also section \ref{vsse}).

The cluster was allowed to evolve dynamically using the adaptive,
block time-step algorithm embedded in \textsc{ph4}.  At regular
intervals of 1 Myr (chosen to resolve mass-loss processes in the most
massive stars, see \citet{PZ99}), the stellar evolution mass loss was
computed and applied to the synchronized N-body model.  \rr{The number
of synchronizations per dynamical time depends upon the dynamical timescale
of the initial model.}  \rr{The simulation was stopped, and the cluster considered
dissolved, if fewer than 12 stars remained in the cluster.}

\begin{figure*}[htpb]
\epsscale{0.5}
\plotone{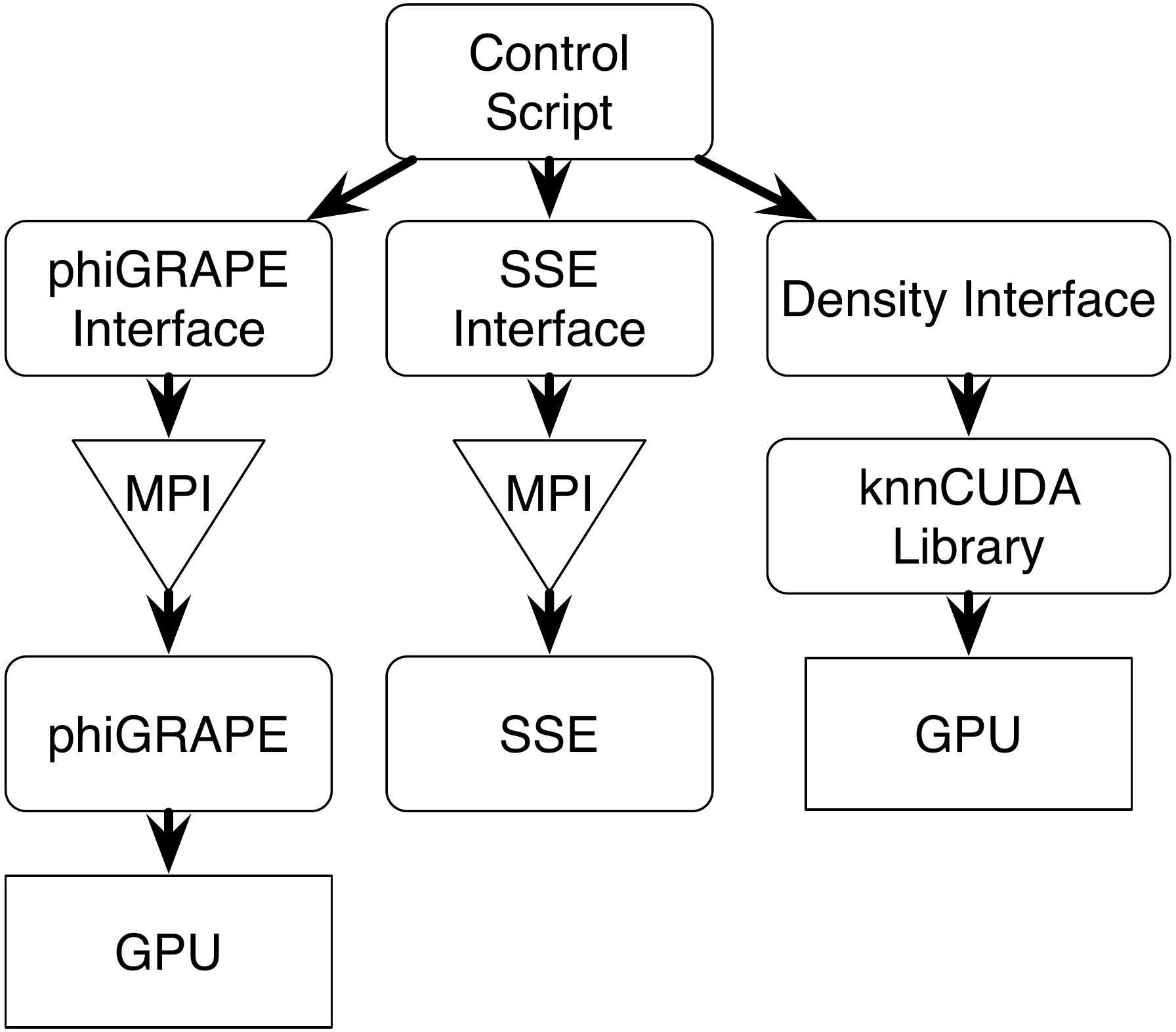}
\epsscale{1}
\caption{The {\AMUSE} software architecture, showing the particular
  modules employed for the bulk of this work and the hardware
  acceleration used.}
\label{archfig}
\end{figure*}

The specific software architecture used in this work is sketched in
Figure \ref{archfig}.  Note that GPU acceleration is used for both
N-body dynamics and for nearest neighbour calculation.

Our runs were conducted on a cluster of 24 dual-socket Intel Xeon X5650, 2.66 GHz nodes with a total of 12 cores each.  Each node contains 64 GB of RAM and six nVidia GTX 480 GPUs.  Our runs ran 
on a single node, using two GPUs: one for dynamics calculation and one for density estimation.  A typical run would use up to three cores: one each for process control, dynamics supervision, and stellar 
evolution.  

\rr{The simulations with \textsc{ph4} running on \textsc{Sapporo}
  using NVidia GPUs performs comparably to GPU-accelerated packages
  such as \textsc{Gadget} 2 and {\Starlab} using \textsc{Sapporo}.  A
  detailed review of the performance of \AMUSE simulations can be
  found in \citet{PZ13}.}  \rr{Energy and angular
  momentum are conserved to within one part in $10^{-7}$ in test runs
  using \textsc{ph4}.  Energy conservation within \AMUSE is studied 
  in more detail in \citet{P13}.}


\section{Physical Model} \label{physmodel}

\subsection{Initial Conditions}

\citet{K66} models with $W_0 = 3$ and $W_0 = 7$ were used as initial
conditions for our model clusters, representing relatively diffuse and
relatively centrally concentrated systems.  \rrtwo{We obtain the King
models by numerical integration, as shown in section III of \citet{K66}.}

A simple power-law stellar
mass function
\begin{equation}
    dN \propto m^{-\alpha} dm
\end{equation}
was used in conjunction with a random number generator to assign
masses to each particle.  Following CW, the slope of the
mass function was taken to be either $\alpha = 1.5$ or $\alpha = 2.5$
with masses ranging from $0.4 M_\Sun$ to $15 M_\Sun$.  A Salpeter mass
function would correspond to $\alpha = 2.35$ \rr{\citep{S55}}.  \rr{For $\alpha = 1.5$,
we expect approximately 15\% of stars to undergo a core collapse supernova.
For $\alpha = 2.5$, the supernova fraction is approximately 2\%.}  \rr{The assumption
of formation by violent relaxation was made.  That is to say that the \rrtwo{initial mass, position and velocity} of a star
\rrtwo{are} uncorrelated, beyond the condition that the cluster begins in virial equilibrium.}

Runs are grouped by ``family,'' a parameter also defined by \rr{CW}.  
The family parameter fixes the relaxation time at the
tidal radius, which effectively changes the ratio of the stellar
evolution and dynamical timescales for a given model.  Four values of
this relaxation time are chosen, as summarized in Table
\ref{familytab}.  The relaxation time \rr{at} the tidal radius is
\begin{equation}
    t_{rlx,CW} = \frac{M^{1/2}r_t^{3/2}}{G^{1/2} m_\star \ln N}
	      = \left(2.57\ \mathrm{Myr}\right) F_{CW}
\end{equation}
where $r_t$ is the King truncation radius and $m_\star$ is a typical stellar mass, here taken to be $M_\Sun$ \rr{(see equation 6 of CW, or equation 8 of TPZ)}.  By making the assumption that $r_t$ is equal to the Jacobi radius $r_J$ of the cluster, 
$F_{CW}$ is defined to be
\begin{equation}
    F_{CW} \equiv \frac{M}{M_\Sun} \frac{R_g}{\mathrm{kpc}} 
		 \frac{220\ \mathrm{km\ s^{-1}}}{v_g} \frac{1}{\ln N}
\end{equation}
where $R_g$ and $v_g$ are the radius and velocity of the cluster, assuming a circular orbit about a parent galaxy (see CW).  In generating our initial conditions, the choice $v_g = 220\ \mathrm{km\ 
s^{-1}}$ is made and $R_g$ is computed to match one of the specified families.

The more conventional \rr{half-mass radius relaxation time} is given by
\begin{eqnarray}
    t_{rh} &=& 0.138 \frac{M^{1/2}r_h^{3/2}}
				{G^{1/2}\left<m\right> \ln N} \nonumber \\*
	   &=& \left(3.54 \times 10^5\ \mathrm{yr}\right) F_{CW}
	    \left( \frac{r_h}{r_t} \right)^{3/2} \frac{M_\Sun}{\left<m\right>}
\label{rheqn}	    
\end{eqnarray}
where $\left<m\right>$ is the average mass of a star in the cluster (TPZ).  For $\alpha = 1.5$ or $2.5$, $\left<m\right> = 2.54 M_\Sun$ or $1.01 M_\Sun$ respectively.  The properties of each family are listed 
in Table \ref{familytab}.  For convenience, a brief summary of the relevant King model parameters is also provided in Table \ref{kingtab}.

\begin{table*}[htpb]
\begin{center}
\caption{Family Summary}
\label{familytab}
\begin{ruledtabular}
\begin{tabular}{cccccccc}
 & &  &  & \multicolumn{4}{c}{$t_{rh}\ (\mathrm{Gyr})$} \\
Family & \rr{$R_g$\footnote{\rr{The value of $R_g$ is slightly different for each random realization, as it is chosen to fix the value of $t_{rlx,CW}$.  A typical value is quoted here.}}} & $F_{CW}$ & $t_{rlx,CW}$ & \multicolumn{2}{c}{$\alpha=1.5$} & \multicolumn{2}{c}{$\alpha=2.5$} \\
 & \rr{($\mathrm{kpc}$)} & & $(\mathrm{Gyr})$ & $W_0=3$ & $W_0=7$ & $W_0 = 3$ & $W_0=7$ \\ \hline
1 & \rr{16.0} & $5.00 \times 10^4$ & 128 & 1.01 & 0.300 & 2.46 & 0.728 \\ \hline
2 & \rr{42.1} & $1.32 \times 10^5$ & 339 & 2.68 & 0.793 & 6.49 & 1.92 \\ \hline
3 & \rr{71.8} & $2.25 \times 10^5$ & 577 & 4.56 & 1.35 & 11.1 & 3.28 \\ \hline
4 & \rr{189} & $5.93 \times 10^5$ & 1522 & 12.0 & 3.56 & 29.2 & 8.64 \\ 
\end{tabular}
\end{ruledtabular}
\end{center}
\end{table*}

\begin{table}[htpb]
\begin{center}
\caption{Relevant King Model Parameters}
\label{kingtab}
\begin{ruledtabular}
\begin{tabular}{cccc}
$W_0$ & $c$ & $r_h/r_t$ & $\rho(0)/\left<\rho\right>$ \\ \hline
3 & 0.67 & 0.27 & 84 \\ \hline
7 & 1.53 & 0.12 & 6430 \\
\end{tabular}
\end{ruledtabular}
\end{center}
\end{table}

The initial conditions were generated using {\Starlab} tools and are
stored in the {\Starlab} format, which is {\AMUSE}-compatible.

\rr{\rrtwo{In this first set of simulations performed with \AMUSE
we have not considered primordial binary stars.}  Further, we
  perform no special treatment of binary or multiple stars in the
  code, as the version of \AMUSE used in most of this work did not
  support such handling.  A softening length of $\epsilon = 1/N$
  (roughly the 90-degree turnaround distance for an equilibrium
  system) was adopted to avoid numerical issues due to close
  encounters and to avoid hard binary formation.  \rrtwo{Taking into account} the fact that our
  models cover only the pre-core collapse regime, during which
  binaries are not expected to form, this omission has no impact on
  our results.  Binary and multiple dynamics are extremely important
  during core collapse and afterward, and will be the subject of Paper
  II in this series, using capabilities subsequently added to
  \AMUSEper}

\subsection{Tidal Truncation}

Tides are simulated by simple truncation at the Jacobi radius
\begin{equation} \label{r_j}
    r_{J} = \left(\frac{M}{3M_{\mathrm{galaxy}}}\right)^{1/3}R_g
\end{equation}
which is assumed to be equal to the King truncation radius $r_t$.  Once every dynamical time any stars which pass beyond the Jacobi radius are removed from the simulation.  The dynamical time for the 
cluster is
\begin{equation}
    t_{dyn} = \frac{GM^{5/2}}{\left(-2U\right)^{3/2}}
    \label{eq:tdyn}
\end{equation}
where $U$ is the gravitational potential energy of the system \citep{PZ10}.


\subsection{Very Simple Stellar Evolution (VSSE)} \label{vsse}

For direct comparison with CW and TPZ, we
implemented the Chernoff \& Weinberg stellar evolution model in
{\AMUSE}.  In this model, a main sequence star remains at constant
mass until it has exceeded its lifetime.  At that point, the star is
transformed into a remnant of lower mass, and evolution ceases.

The stellar lifetime is determined by fitting a cubic spline to the data points listed in Table \ref{lifetimetab}, which were taken by Chernoff \& Weinberg from \citet{MS79}.  The remnant mass is derived using 
Equation \ref{remnantmass}, which was based on \citet{IR83}.  Note that the lifetimes listed are assumed to correspond to Population I stars.

\begin{table}[htpb]
\begin{center}
\caption{VSSE Lifetimes}
\begin{ruledtabular}
\label{lifetimetab}
\begin{tabular}{cc}
Initial Mass & Main Sequence Lifetime \\
($\log_{10} \left[m_i/M_\Sun\right]$) & ($\log_{10}\left[t_{se}/\mathrm{yr}\right]$) \\ \hline
1.79 & 6.50 \\
1.55 & 6.57 \\
1.33 & 6.76 \\
1.11 & 7.02 \\
0.91 & 7.33 \\
0.72 & 7.68 \\
0.54 & 8.11 \\
0.40 & 8.50 \\
0.27 & 8.90 \\
0.16 & 9.28 \\
0.07 & 9.63 \\
-0.01 & 9.93 \\
-0.08 & 10.18 \\
\end{tabular}
\end{ruledtabular}
\end{center}
\end{table}

\begin{equation}
\label{remnantmass}
m_f = \begin{cases}
0.58 M_\Sun + 0.22 \left(m_i - M_\Sun \right), & m_i < 4.7 M_\Sun \\
0.0, & 4.7M_\Sun \le m_i \le 8.0 M_\Sun \\
1.4 M_\Sun, & 8.0 M_\Sun < m_i \le 15.0 M_\Sun \\
\end{cases}
\end{equation}


\section{Results and Discussion} \label{resultssec}

\subsection{Validation}\label{validationsec}

\begin{figure*}[htpb]
\includegraphics[width=3.5in]{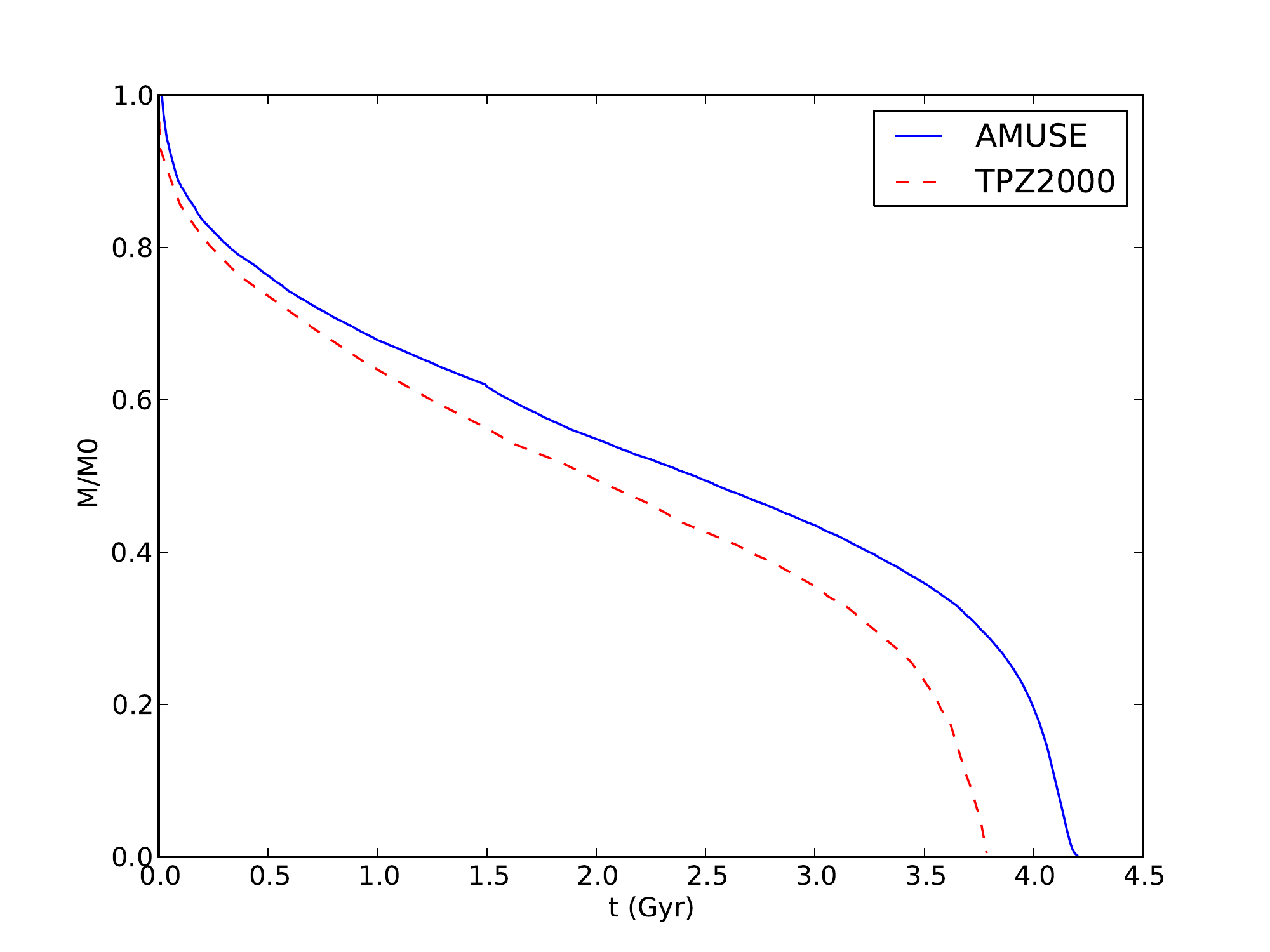}\includegraphics[width=3.5in]{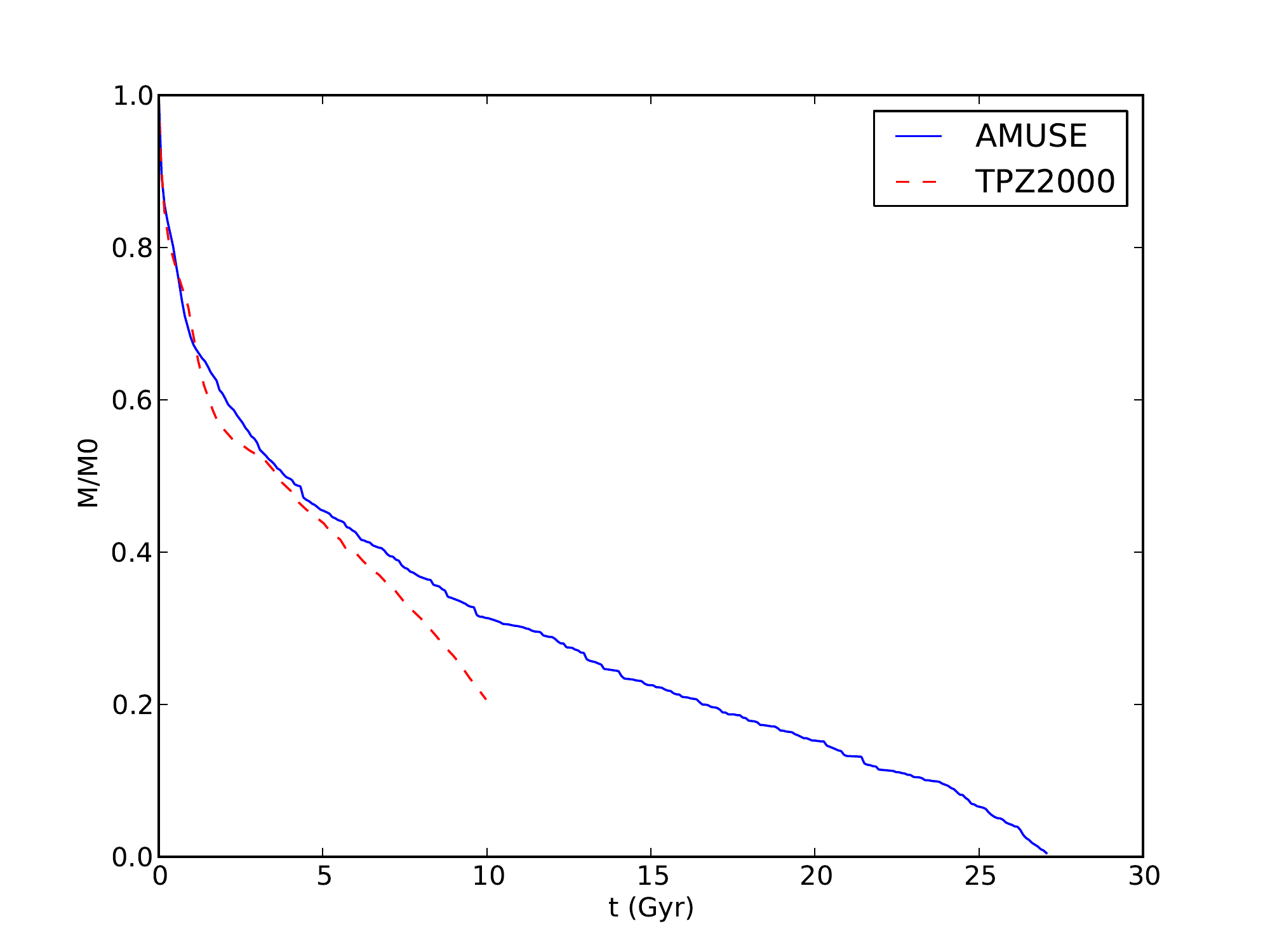}
\includegraphics[width=3.5in]{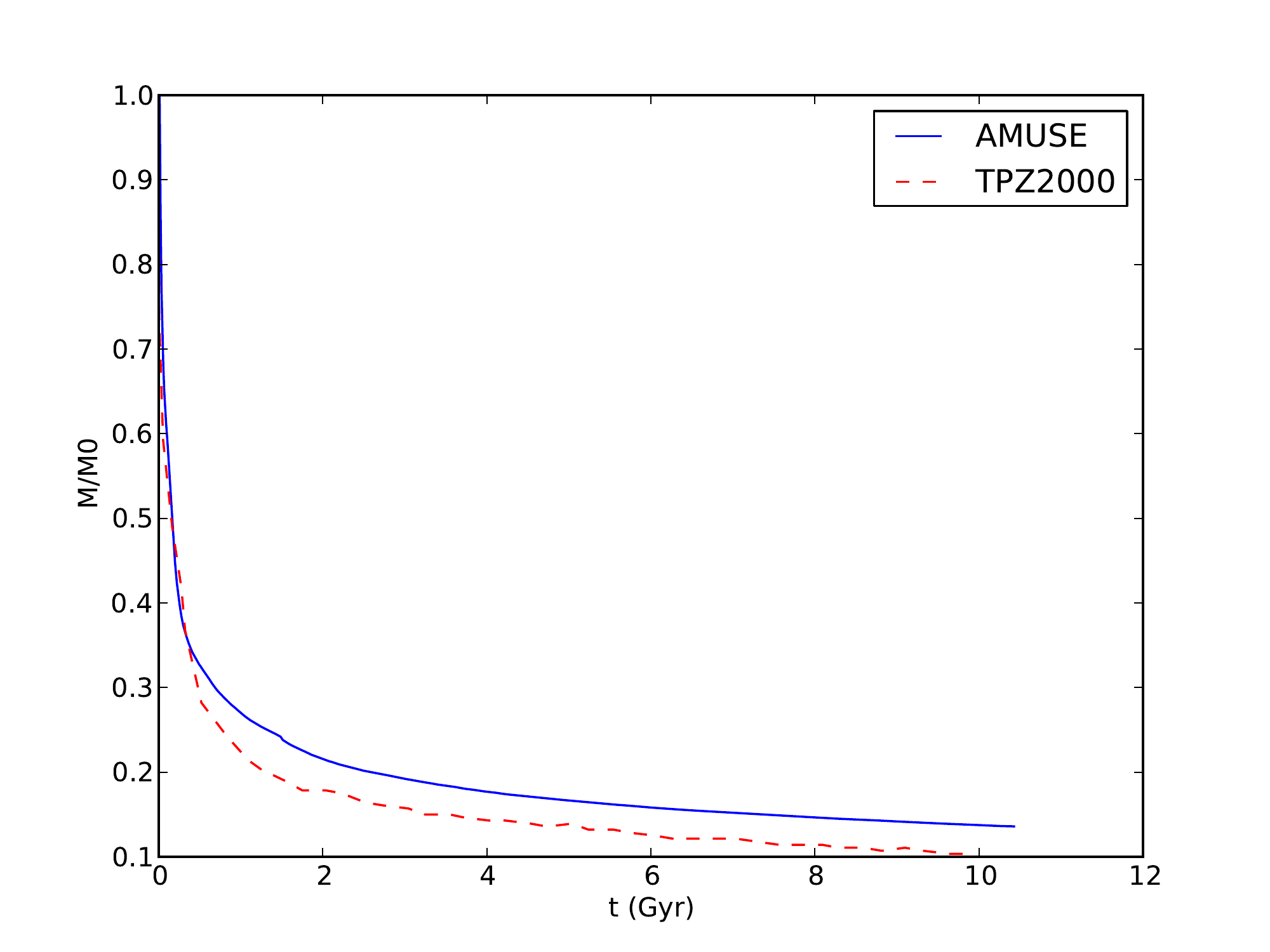}\includegraphics[width=3.5in]{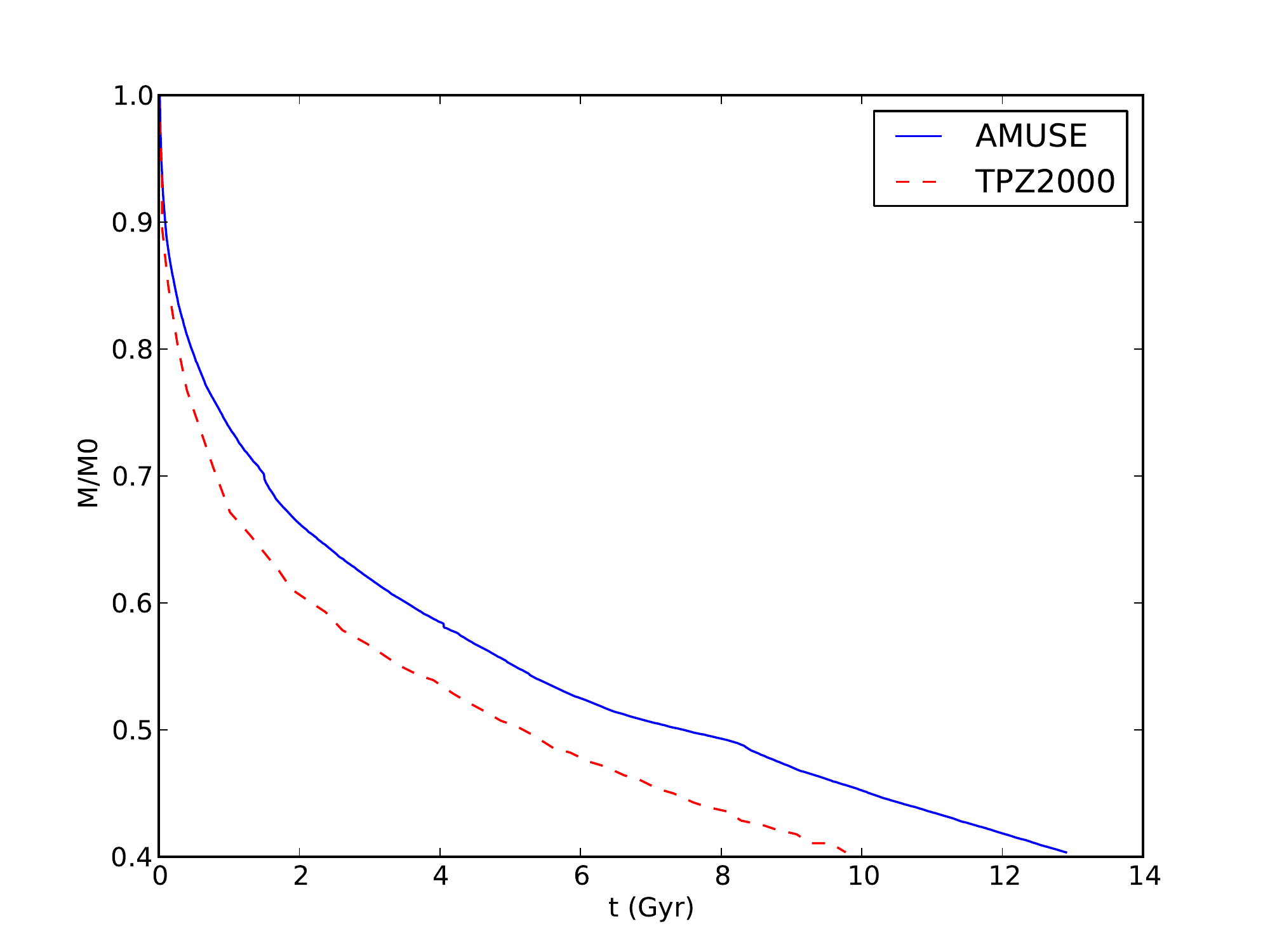}
\caption{Evolution of the mass of a cluster of N=32,000 stars, with
  initial cluster parameters ($W_0, \alpha, \mathrm{family}$), from
  left to right: (3, 2.5, 1), (3, 2.5, 4), (7,1.5, 4) and (7, 2.5, 1).
  The track produced by {\AMUSE} is shown as the solid line (blue in
  the electronic version).  The corresponding run result from
  \citet{TPZ00} (N-body model) is shown as the
  dashed line (red in the electronic version).}
\label{valfig}
\end{figure*}

We first set out to test whether or not the AMUSE framework could
reproduce known results.  CW produced a well studied catalog
of Fokker--Planck code simulations.  TPZ performed runs for
the same initial parameters using both a Fokker--Planck code and an
N-body code, calibrating the removal of escapers in the former against
the latter.  Our initial goal was to see if, and how well, AMUSE could
replicate the TPZ N-body results.

In TPZ, curves for four choices of parameters are published.  Using the
notation $(W_0, \alpha, \mathrm{family})$, they are: (3, 2.5, 1), (3, 2.5, 4), 
(7, 1.5, 4) and (7, 2.5, 1).  We therefore generated initial conditions for each
of these cases and ran them using \AMUSEper

Figure \ref{valfig} shows the result of our attempt to reproduce
selected TPZ runs.  The qualitative shape of the curves are
similar, but it is difficult to make any quantitative statement about
the agreement or disagreement between the two sets of simulations.

An obvious suspect for these variances is the difference in stellar
evolution recipes, which are known to be similar but not identical.
TPZ used SeBa \citep{PZV96} to model stellar evolution, while we have
used SSE.  It is likely that small differences in the handling of main
sequence lifetime and remnant masses for short-lived O and B stars
could alter the later evolution of the cluster due to early mass loss.
This is not the only possibility.  \S\ref{secomp} reviews the variance
introduced by the choice of stellar evolution model.

One known difference is in the handling of kicks in supernova formation.
Our top-level \AMUSE script ignores any kick prescriptions built into the
stellar evolution codes.  However, some of the models (for example, CW and SSE)
leave a zero-mass ``remnant'' for some initial stellar masses.  This represents
the detonation of the remnant during the supernova, and the ejection of any remaining
material.  Our \AMUSE script treats this as the star disappearing.  Slight differences 
in these prescriptions are likely to contribute to variations seen between stellar evolution codes.
It is unlikely, however, that this was the dominant source of variation.

This led us to consider
the inherent spread in our results due to random variations in the
initial conditions (individual star masses, positions, and
velocities), as opposed to systematic differences between the
integrators.  We conducted multiple random realizations of each of our chosen initial parameter sets.  Each parameter set (for example, $W_0=3, \alpha=2.5, \mathrm{\rrtwo{family}}=1$) describes a continuous model, and we transform this model into a discrete mass spectrum and set of spatial and velocity data using a variable random number seed.  We refer to a set of random realizations of a given parameter set as ``formally equivalent.''

\begin{figure*}[htpb] 
\plottwo{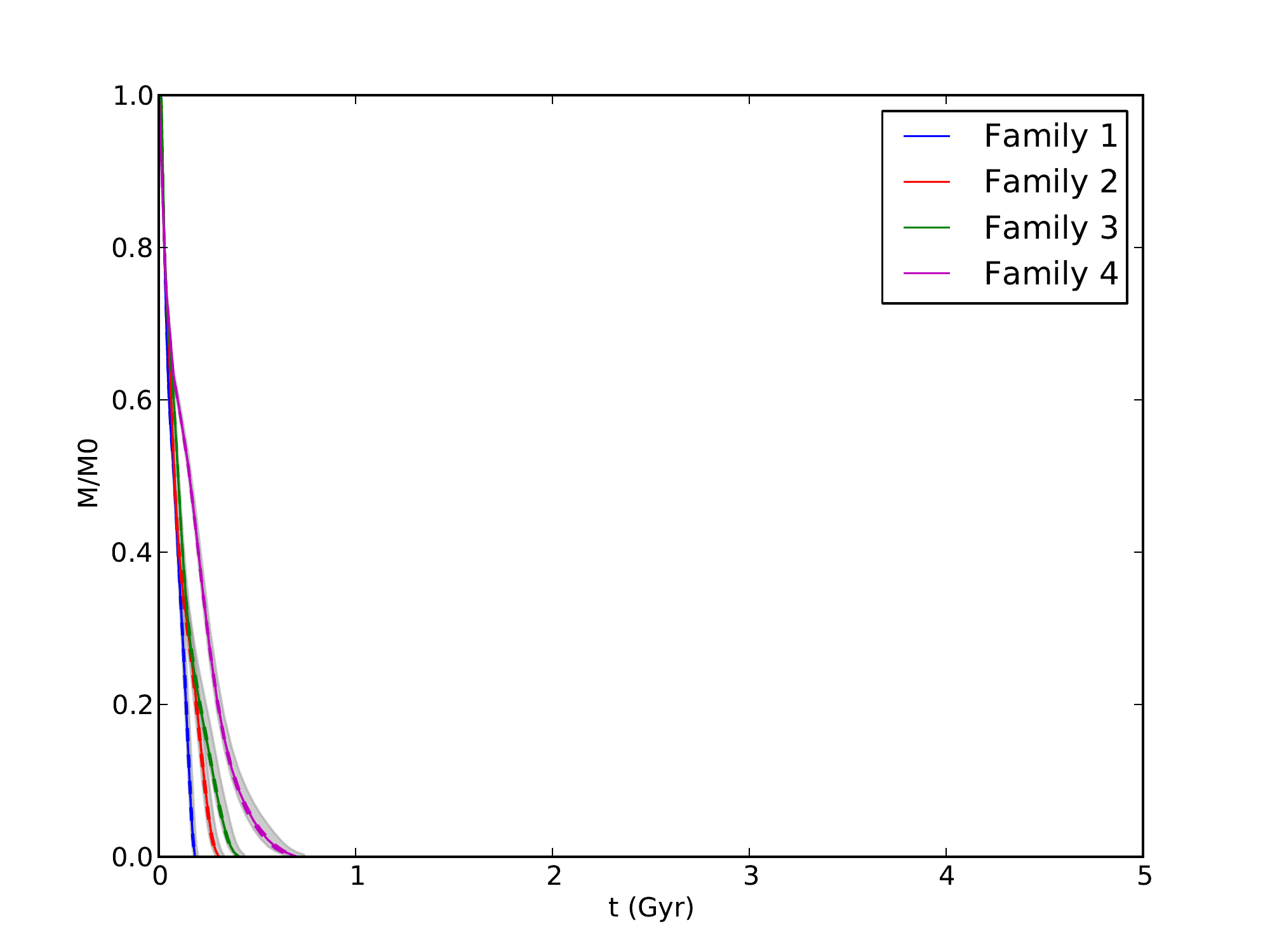}{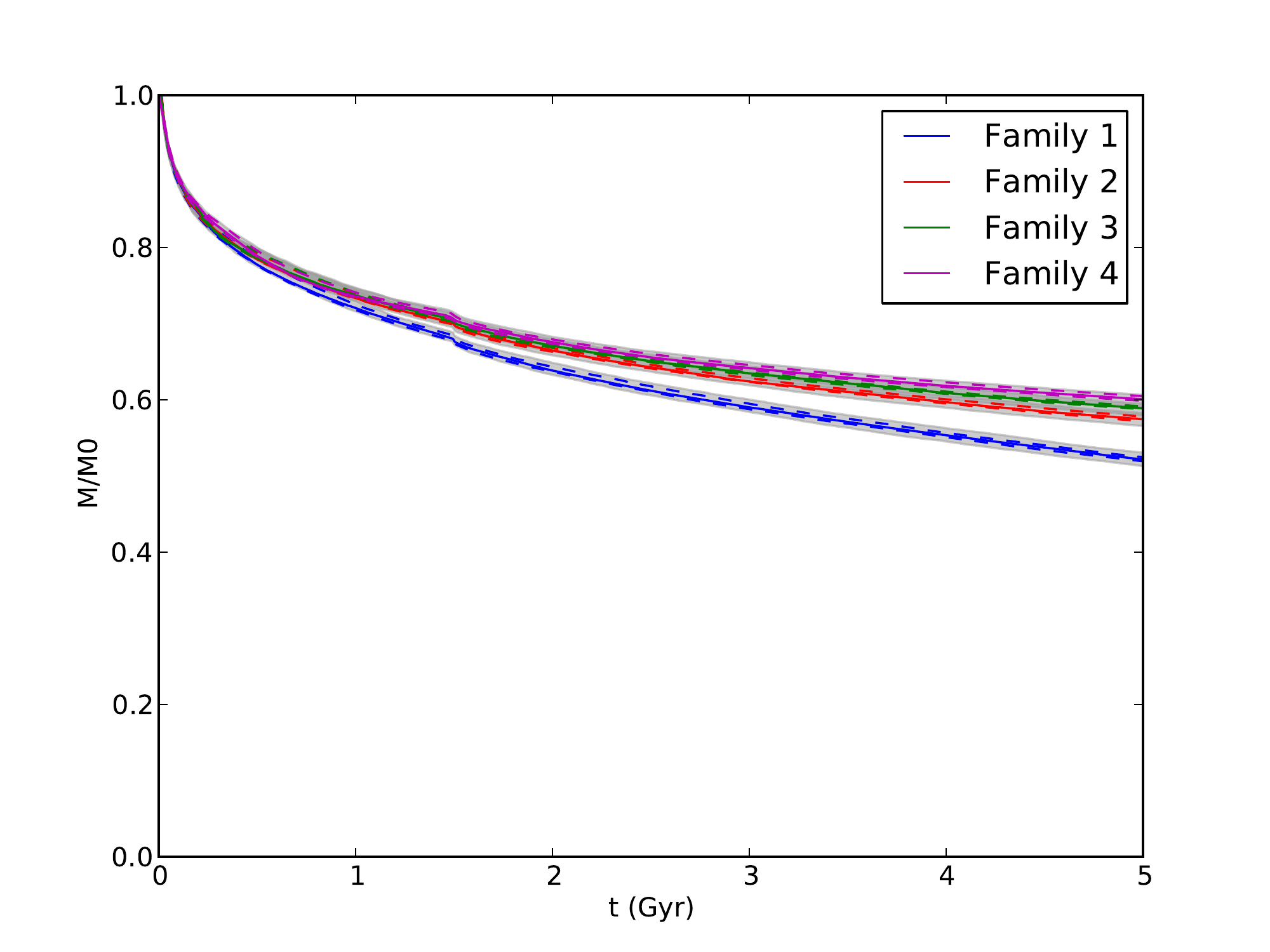}\\
\epsscale{0.9}
\plotone{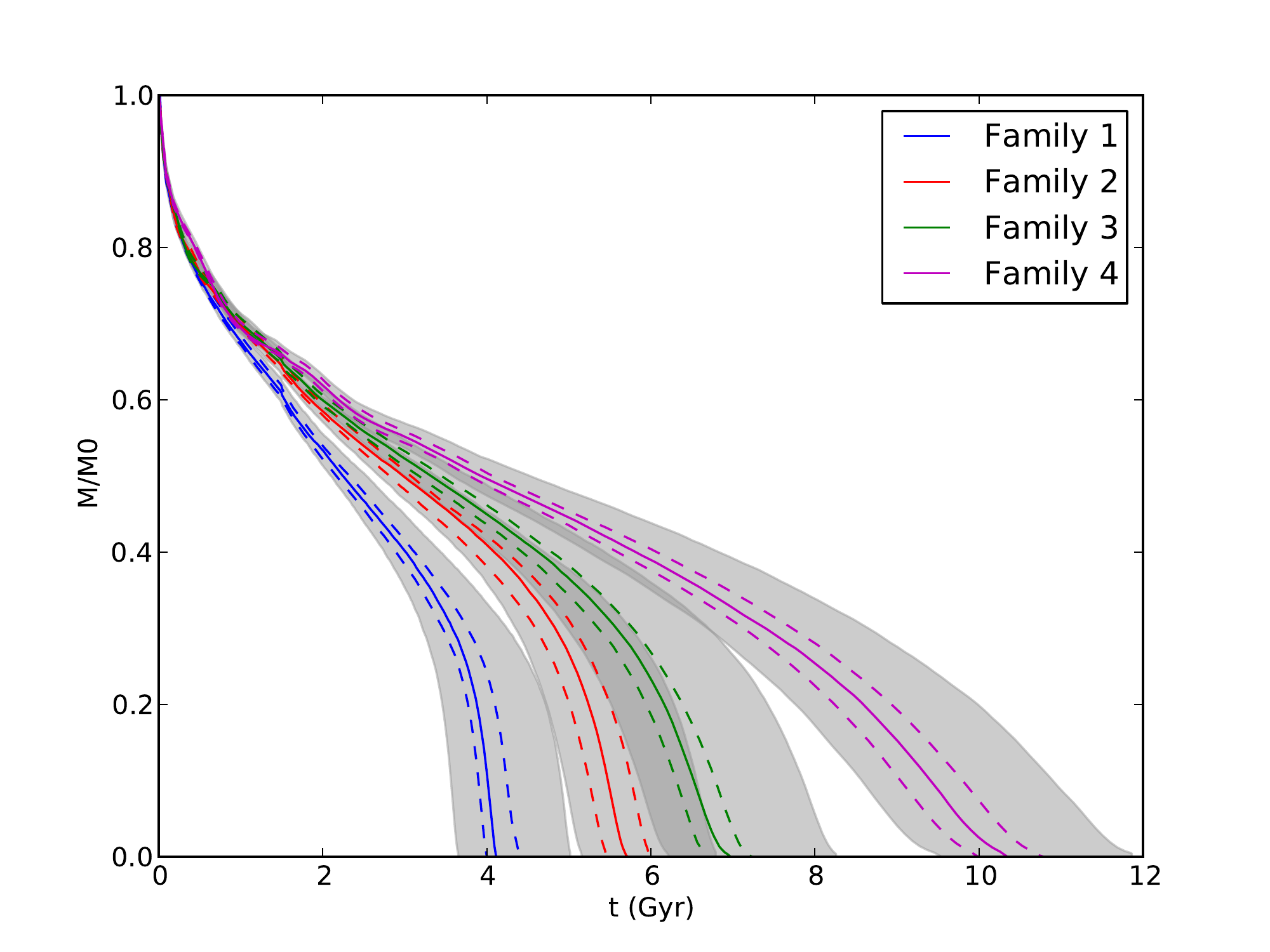}
\epsscale{1.0}
\caption{Evolution of the mass of a cluster of N=32,000 stars, with initial cluster parameters ($W_0, \alpha$), from top-left clockwise: (3, 1.5), (7, 2.5), and (3, 2.5).  For each family, the solid line indicates the 
median value, the dashed lines indicated the $25^{th}$ and $75^{th}$ percentiles, and the shaded region indicates the total parameter space explored.  This visualization is related to that in \citet{E11}.}
\label{21runplot}
\end{figure*}

Figure \ref{21runplot} shows the results of 21 different realizations
of three selected models.  These models are the $(W_0, \alpha)$ 
combinations (3, 1.5), (7, 2.5) and (3, 2.5) with all four families simulated
for each case.  This subset is 12 of the 16 possible combinations of
$W_0$, $\alpha$ and family for the TPZ choices of these parameters.
We do not show (7, 1.5) as it is largely uninteresting: the 
cluster dissolves quickly, much like the (3, 1.5) case.  Furthermore, this
is not a useful parameter set for validation because TPZ do not publish
their cluster mass evolution curves for these combinations of parameters.

The variation due to randomness in the initial conditions is small,
except for the case $W_0 = 3, \alpha=2.5$, where the cluster
dissolves, but ``slowly'' relative to stellar evolution.  ``Slowly''
refers to the fact that the initial mass loss driven by supernovae of
O and B stars is not sufficient to disrupt the cluster.  We explore this in further detail in \S\ref{sec:typesofdis}. 
For $W_0 = 3, \alpha=1.5$, the early
stellar evolution loss due to massive stars dominates the cluster's
evolution, and it dissolves before differences in initial conditions
can have much effect.  Conversely, for $W_0=7, \alpha=2.5$, the
cluster is quite long-lived and the effects of initial condition
variability are smoothed out over time.

\subsection{Comparison with TPZ}\label{sec:comparisontpz}

\begin{figure*}[htpb] 
\epsscale{1.17}
\plottwo{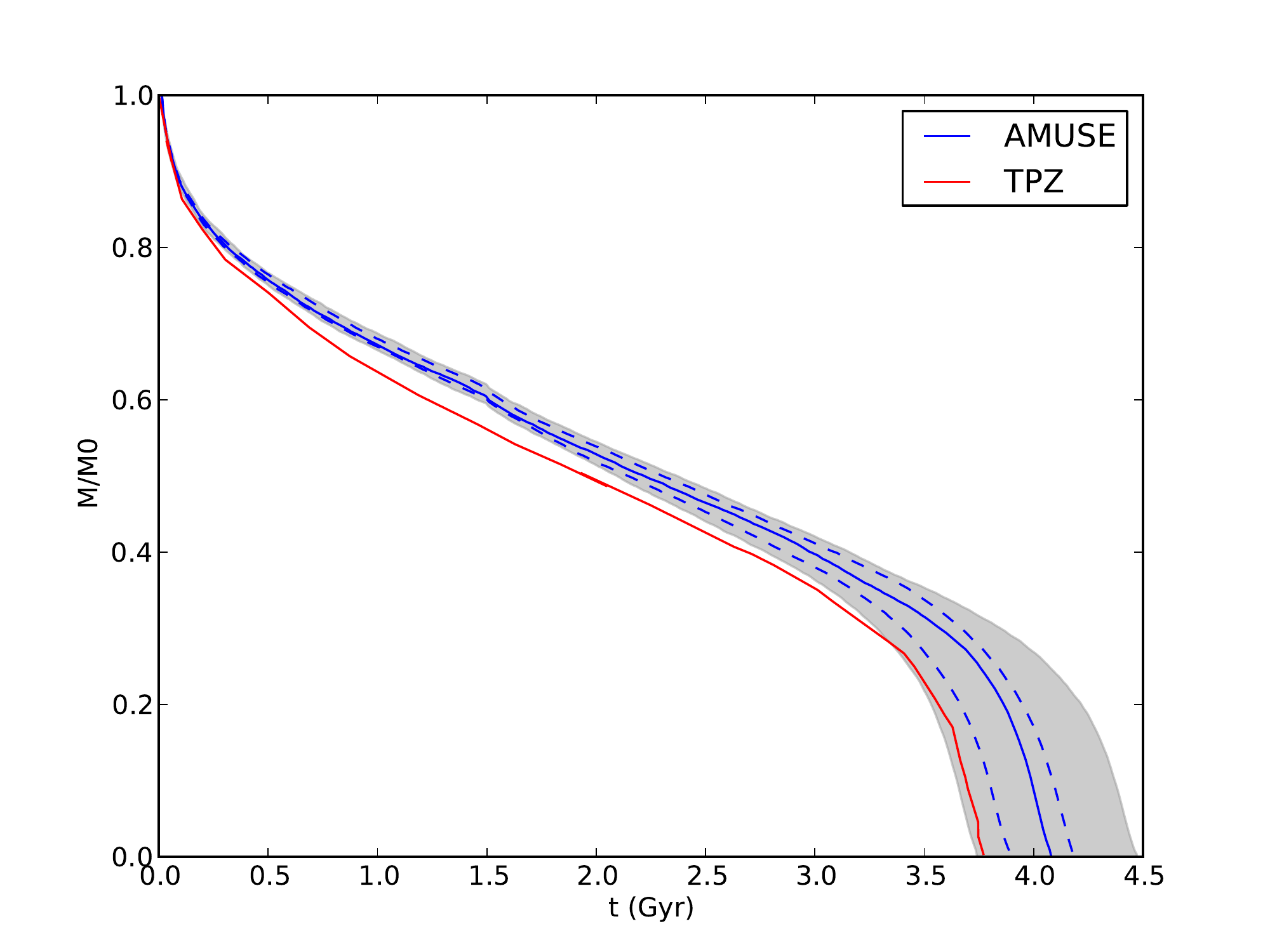}{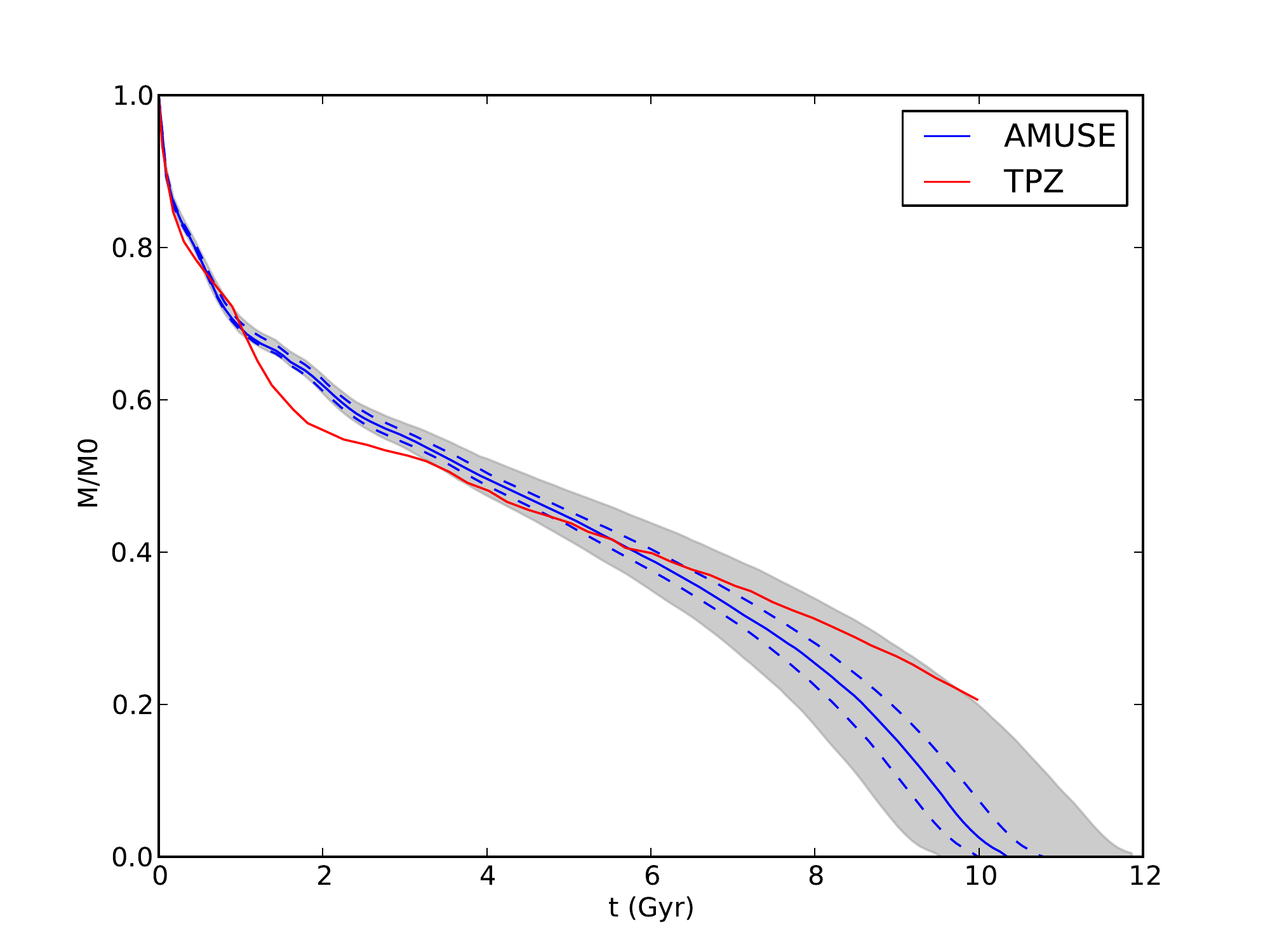}
\caption{Evolution of the mass of a cluster of N=32,000 stars, with initial cluster parameters $W_0 = 3$, $\alpha = 2.5$, $\mathrm{family} = 1$ (left) or $\mathrm{family} = 4$ (right).  The range of variation 
in the {\AMUSE} runs (made using \textsc{ph4} and \textsc{SeBa}) is shown, with solid lines indicating the median, dashed lines the $25^\mathrm{th}$ and $75^\mathrm{th}$ percentiles, and the shaded area the entire envelope of the 
explored parameter space.  A comparison is made to runs published by \citet{TPZ00} in red.}
\label{tpzcompplot}
\end{figure*}

\begin{figure*}[htpb] 
\epsscale{0.5}
\plotone{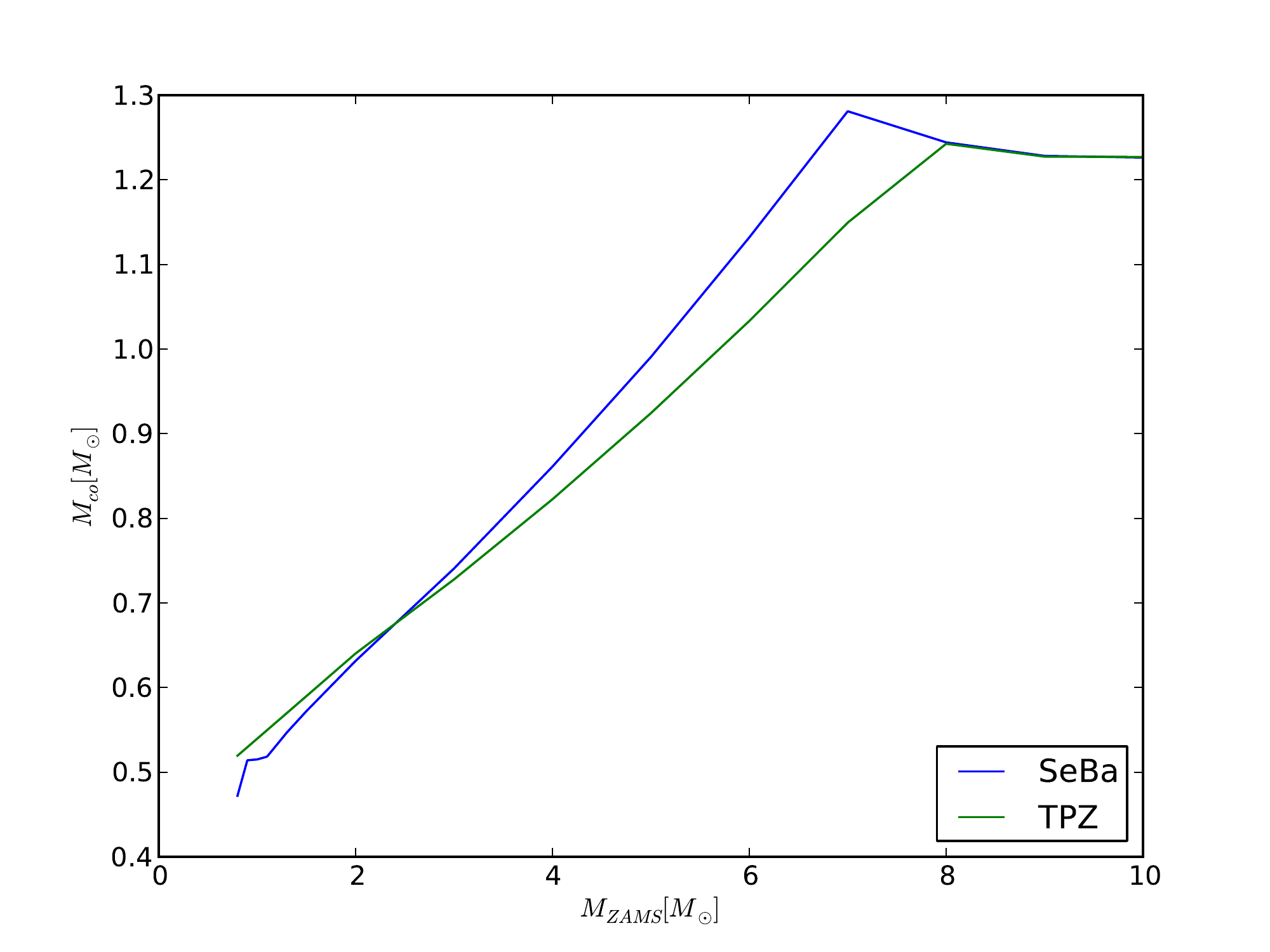}
\caption{Change in the handling of remnants within \textsc{SeBa} between TPZ (published 2000) 
and the {\AMUSE} \textsc{SeBa} module.  Note that \textsc{SeBa} has been improved to better match
the current understanding of stellar evolution.}
\label{sebadrift}
\end{figure*}

Figure \ref{tpzcompplot} also shows curves derived from TPZ's \Starlab runs 
overplotted on an {\AMUSE} simulation made using 21 runs done with \textsc{ph4}
and the \textsc{SeBa} module.  Despite using the ``same'' stellar evolution recipe,
the results clearly do not agree.  This is due to ``drift'' in the \textsc{SeBa} code since
the TPZ paper was published in 2000.  ``Drift'' refers to changes in the code underlying \textsc{SeBa}
over time as the stellar evolution model is improved.  Figure \ref{sebadrift} shows one such change:
the remnant mass kept by the code has changed in the 12 \rrtwo{yr} between the publication
of TPZ and the current \textsc{SeBa}.

The lesson to simulators is clear: codes can change with time, and these changes can
produce significant differences in the results of simulations.

\subsection{Sources of Variance}
\label{sec:sources}

We would like to know why some of the curves in Fig. \ref{21runplot} have
such large spreads in their tracks while others are constrained to a narrow
area.  In particular, the $W_0=3; \alpha=2.5$ case behaves quite differently
from the other cases.  There are two obvious
places in the discretization process where random variations might
play a decisive role---the masses of the most massive stars, which
explode early and can eject a significant amount of material from the
cluster, and their locations, since mass ejected from the cluster
center is much more destructive to the cluster than mass ejected from
the outer regions \citep{V09,P12}.

\begin{figure*}[htpb] 
\plottwo{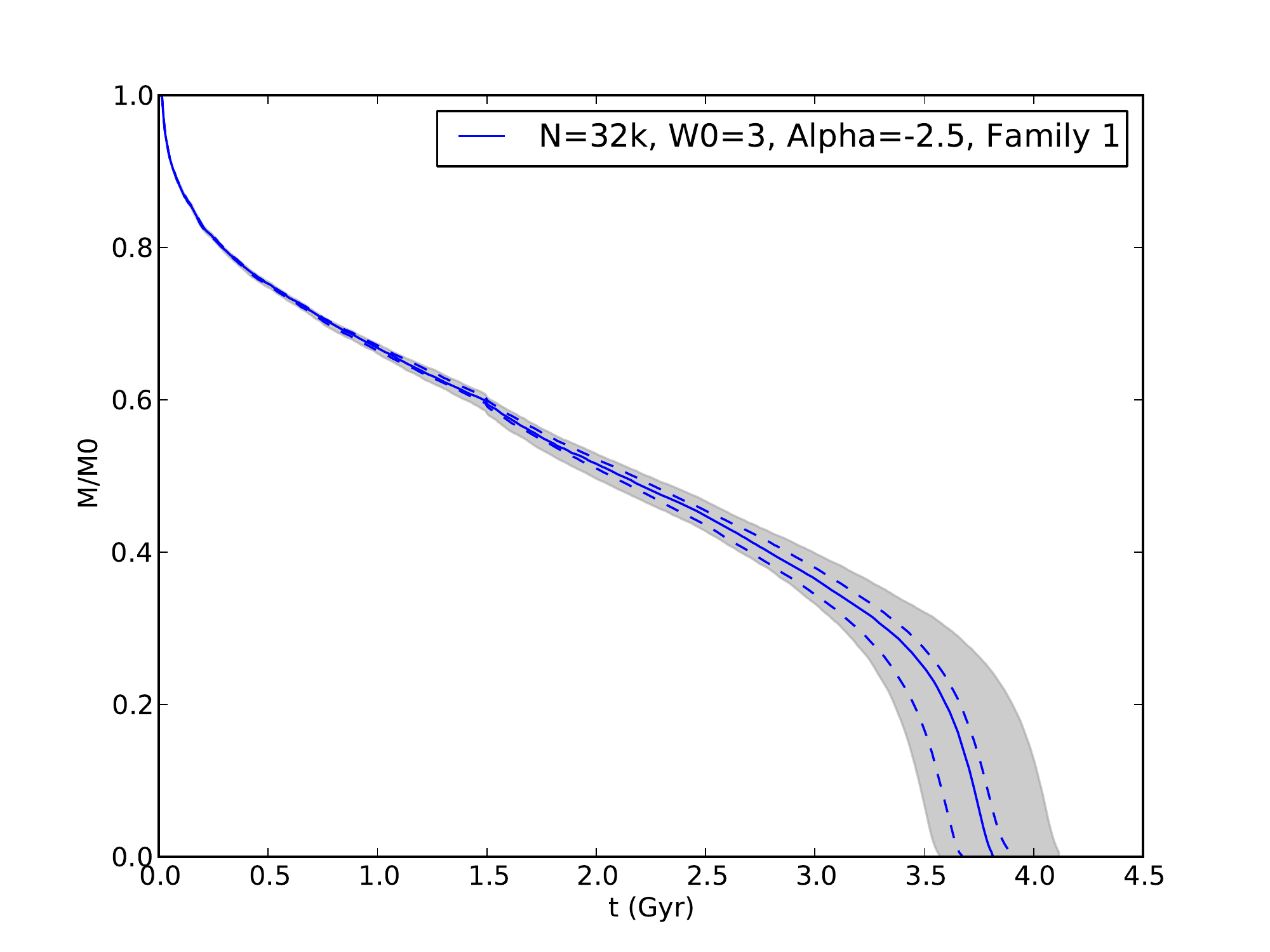}{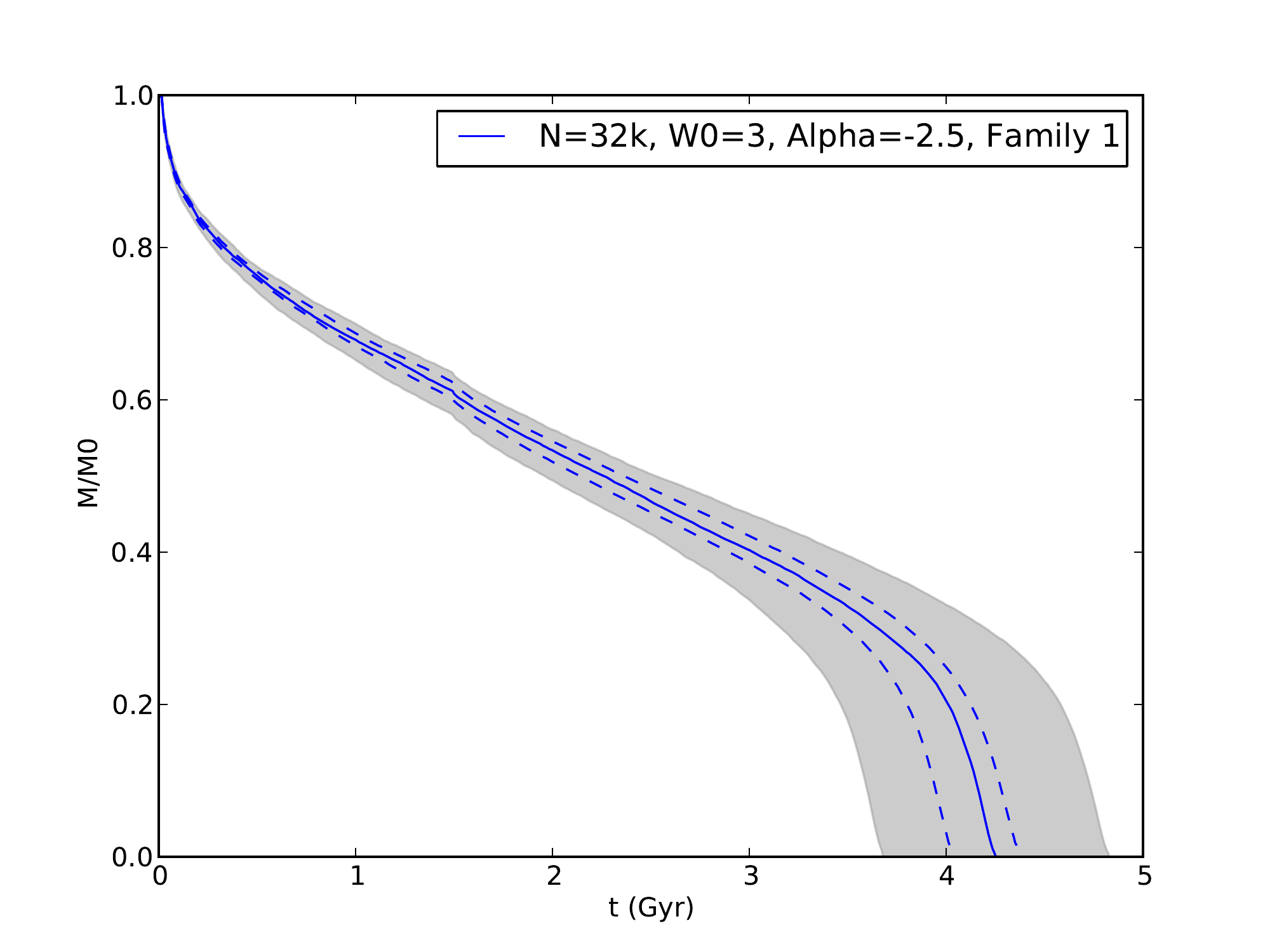}
\caption{Evolution of the mass of a cluster of N=32,000 stars, with
  initial cluster parameters $W_0 = 3$, $\alpha = 2.5$,
  $\mathrm{family} = 1$.  The range of variation seen in {\AMUSE} runs
  is shown in blue with solid indicating the median, dashed lines the
  $25^\mathrm{th}$ and $75^\mathrm{th}$ percentiles, and the shaded
  area the entire envelope of explored parameter space.  The left-hand
  plot shows the result when the same mass list is applied to
  differing spatial positions, while the right-hand plot shows the
  result when the same spatial positions are matched to differing mass
  lists.}
\label{splitfig}
\end{figure*}

To separate these two effects, we explored the consequences of holding
one of these ``random'' parameters constant and varying the other.  For
each of the two parameters we ran 11 simulations with random choices
of the other.  Figure \ref{splitfig} illustrates the results.  It is
clear from the plot that varying the masses while holding the positions fixed
has a larger effect than holding the masses constant and varying the positions.  
The mass spectrum effect is larger than the
spatial effect by a factor of about 2.  These two effects are of the same 
order, and neither is negligible when comparing simulation results.

\subsection{Types of Dissolution} \label{sec:typesofdis}


\rr{We now focus our attention on those systems whose dissolution
  timescale is significantly shorter than that due solely to
  relaxation-driven mass loss, but longer than that due to early
  stellar-evolution mass loss.  These clusters' lifetimes may be as
  long as several \rrtwo{Gyr}, but their dissolution should still be
  considered dynamical. This behaviour was also observed by
  \citet{CW90,F95,TPZ00}.  \citet{F95} explained the mechanism of
  final disruption as a result of the loss of dynamical equilibrium
  within the cluster.  We have examined the boundary between the
  ratios of the various timescales involved, in order to map the
  boundary separating the slow and rapidly dissolving regimes.}

There are three competing timescales in this simulation: the stellar mass loss \rrtwo{timescale}
 $t_{SE}$ (which is of order a few tens of Myr at the beginning of the simulation, and 
scales proportionally to $t$ as the evolution of the cluster progresses), the
dynamical timescale $t_{dyn}$ (defined in equation \ref{eq:tdyn}) and the relaxation \rrtwo{timescale}
 $t_{rh}$ (defined in equation \ref{rheqn}).  For a given choice
of mass function slope $\alpha$ the stellar mass loss timescale is fixed; it is an intrinsic property 
of the stars themselves and not of their dynamical phase space configuration. 

The runs shown in Fig. \ref{21runplot} show two modes of cluster dissolution:
the flat slope of the long-lived ($W_0=7$, $\alpha=2.5$) runs or the short-lived
``ski jump''  curve of the short-lived ($W_0=3$, $\alpha=2.5$) runs.  The ($W_0=3$, $\alpha=1.5$) 
runs live extremely short lives.  Nevertheless, it is clear from examining the family 4 curve
that the ``ski jump''  is present in them too, and that they are more similar
to ($W_0=3$, $\alpha=2.5$) than to ($W_0=7$, $\alpha=2.5$).  It is clear that the
lifecycle of massive stars plays a role in the disruption of the cluster in the short-lived
``ski jump'' cases.  In the long-lived case of ($W_0=7$, $\alpha=2.5$), the cluster
survives the early stellar mass loss and evolves according to relaxation processes.

\begin{figure*}[htpb] 
\epsscale{0.9}
\plotone{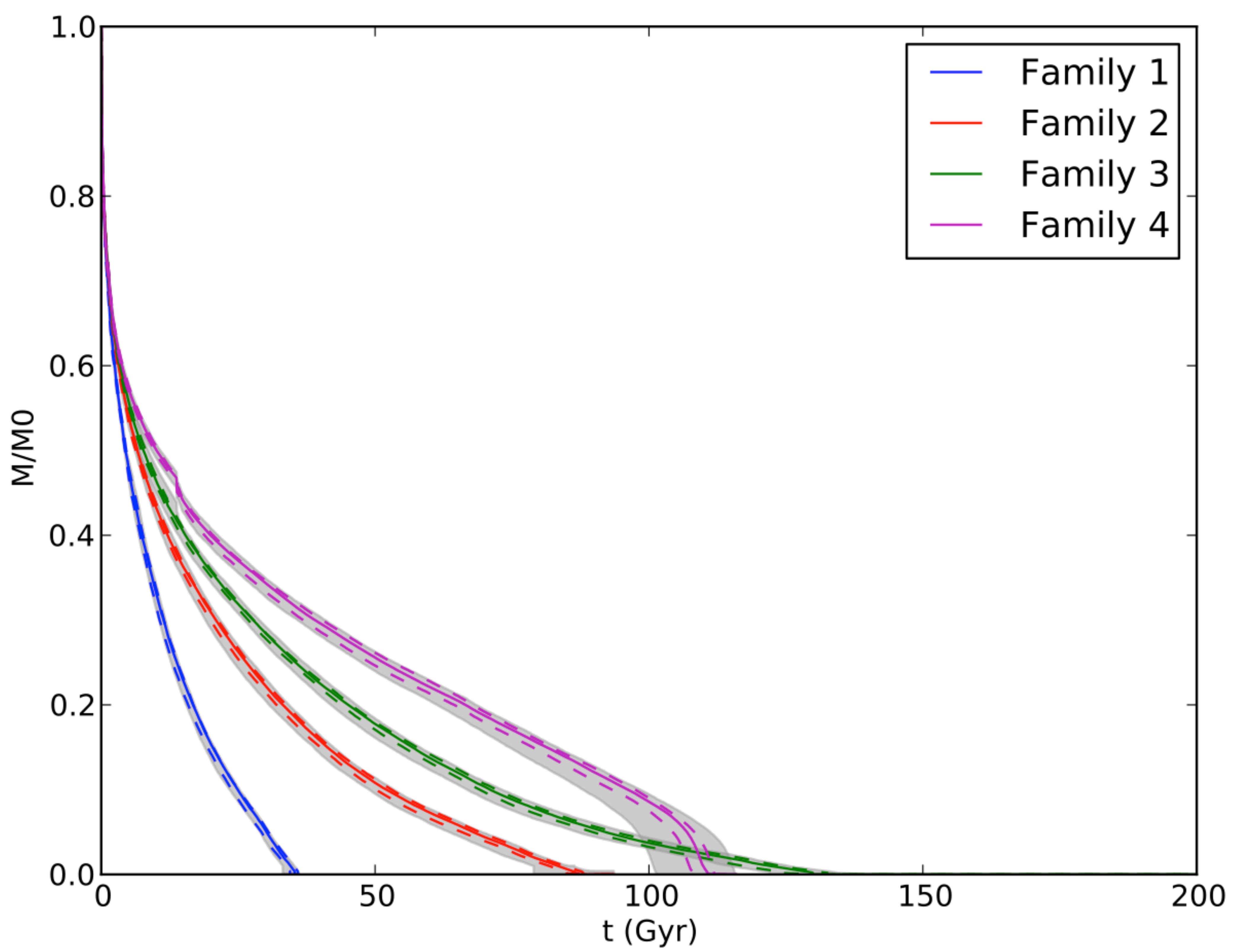}
\epsscale{1.0}
\plottwo{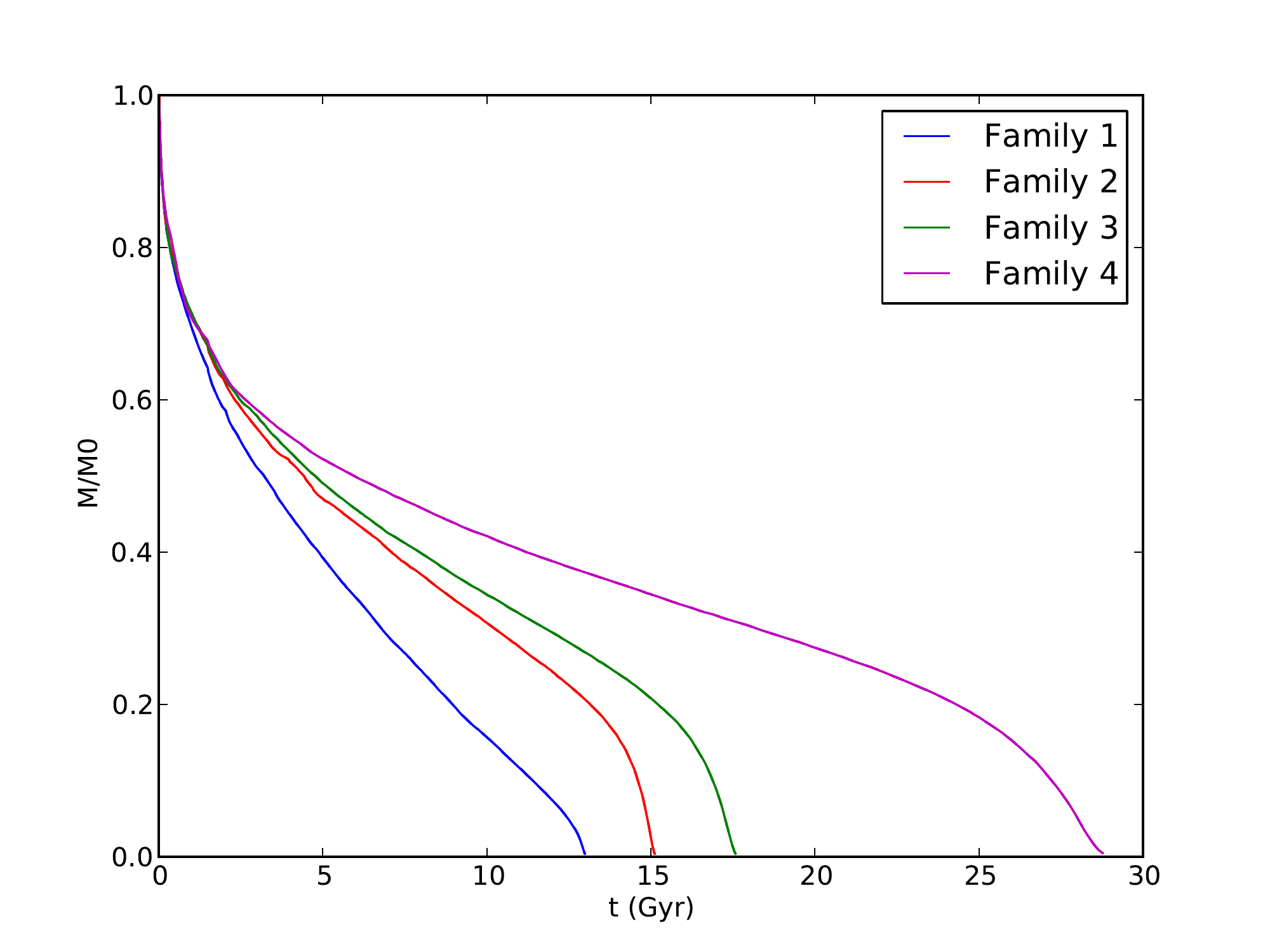}{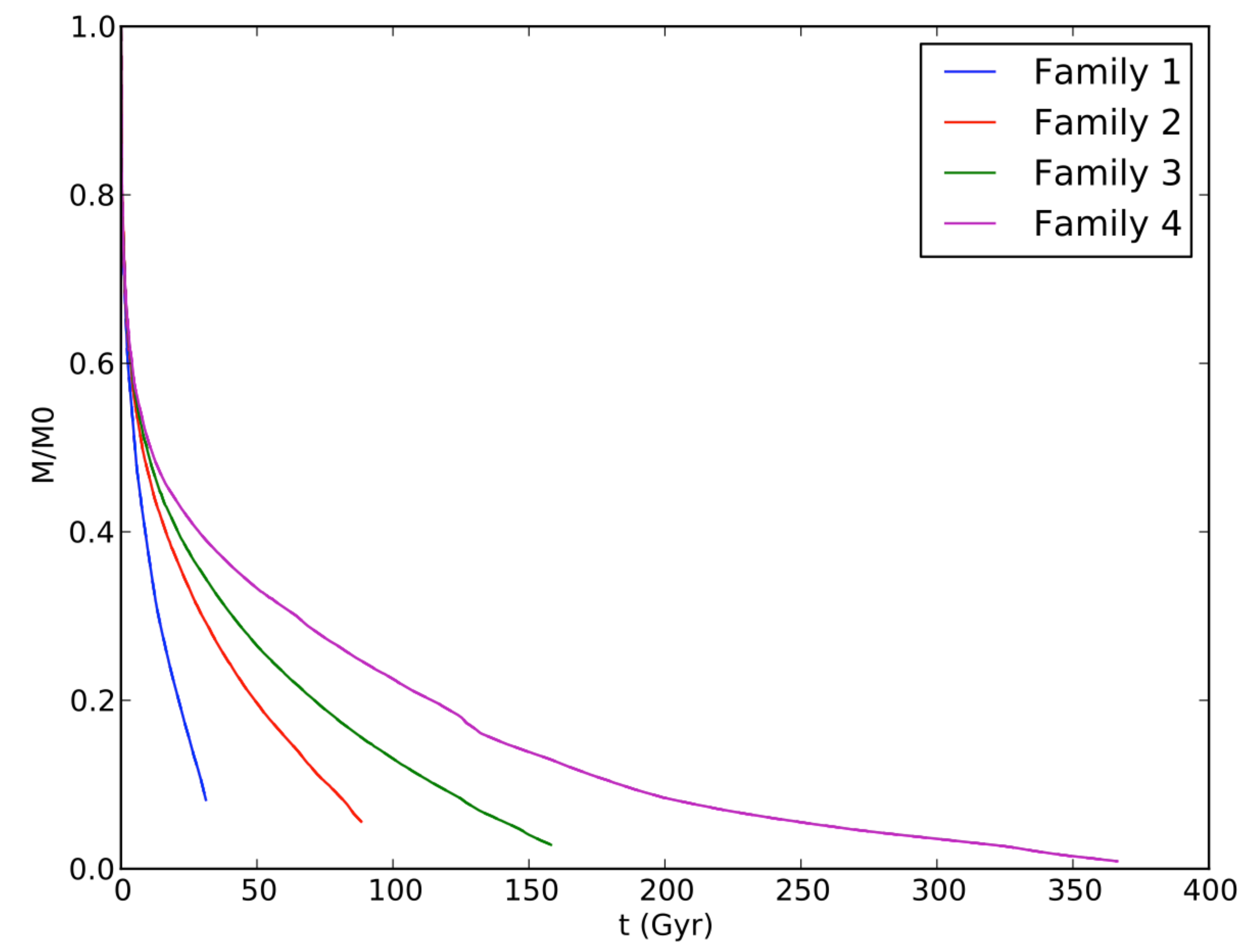}\\
\caption{Evolution of the mass of a cluster of N=32,000 stars, with initial cluster parameters ($W_0, \alpha$), from top-left clockwise: (4, 2.5), (6, 2.5), and (5, 2.5).  For each family, the solid line indicates the 
median value, the dashed lines indicated the $25^{th}$ and $75^{th}$ percentiles, and the shaded region indicates the total parameter space explored.  The families are 1, 2, 3, and 4 in order from left to right.  Note that family 4 in the $W_0=5$ case shows dynamical dissolution behaviour while the other three families show relaxation dissolution behaviour.}
\label{additionalruns}
\end{figure*}

This inspired additional runs, conducted for $W_0=4$, $5$, and $6$ with $\alpha=2.5$.
A randomly-realized selection of 10 runs was conducted for each family, and the results
are plotted in Fig. \ref{additionalruns}.  The case ($W_0=5$, $\alpha=2.5$, \rrtwo{family}=$4$) 
is particularly interesting since it dissolves via dynamical processes while the remainder
of the ($W_0=5$, $\alpha=2.5$) models dissolve via relaxation processes.

\begin{figure*}[htpb]
\epsscale{0.8}
\plotone{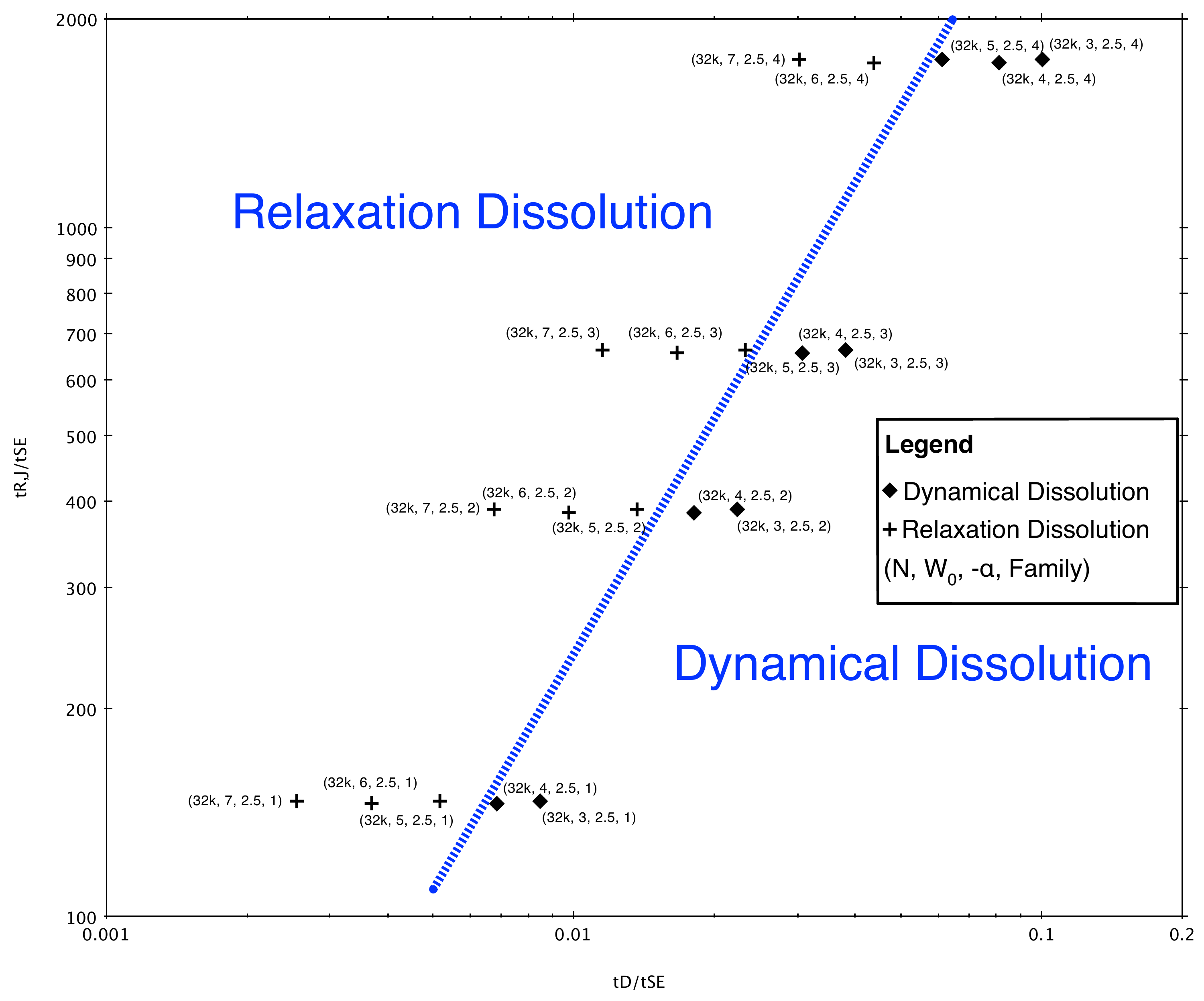}
\caption{The space of combinations of the three fundamental star cluster timescales for $N=32k$.  We 
identify two permitted regions where the death of the star cluster can be characterized as 
due to ``disruption'' (cluster dissolution triggered by the death of massive stars) or due to ``relaxation'' 
effects (due to the slow evolution of the cluster's dynamical parameters).  The label for each point is
($N$, $W_0$, $\alpha$, \rrtwo{family}).  Dissolution results are denoted with diamonds.  Relaxation results
are shown as crosses.}
\label{theofig}
\end{figure*}

Figure \ref{theofig} is a schematic of the relationship between the three timescales for $N=32k$.
We have subdivided the permitted region into two areas.
In the right-most region, the stellar mass loss timescale is much smaller than the dynamical \rrtwo{timescale}.
  This corresponds to the behaviour seen in the ($W_0=3$, $\alpha=1.5$) runs:
the cluster is rapidly disrupted by the death of massive stars early in its life.  In this region,
clusters dissolve due to disruption.

In the left-most region, the opposite is true.  In
the left-most region as the ratio of the dynamical to stellar mass
loss timescales decreases so do the effects of stellar mass loss on the
cluster structural evolution. For systems in this region (e.g. the
family of models with ($W_0=7$, $\alpha=2.5$)), the cluster dissolution is driven
mainly by two-body relaxation mass loss.  In our terminology, the cluster dissolves due to relaxation.

There is a small forbidden region near the horizontal axis which is not shown due to its small size.
This region is forbidden because $t_{rlx,CW}$ cannot be smaller than $t_{dyn}$.  Similar diagrams exist for
other values of $N$ and $\alpha$.

\begin{figure*}[htpb]
\epsscale{0.8}
\plotone{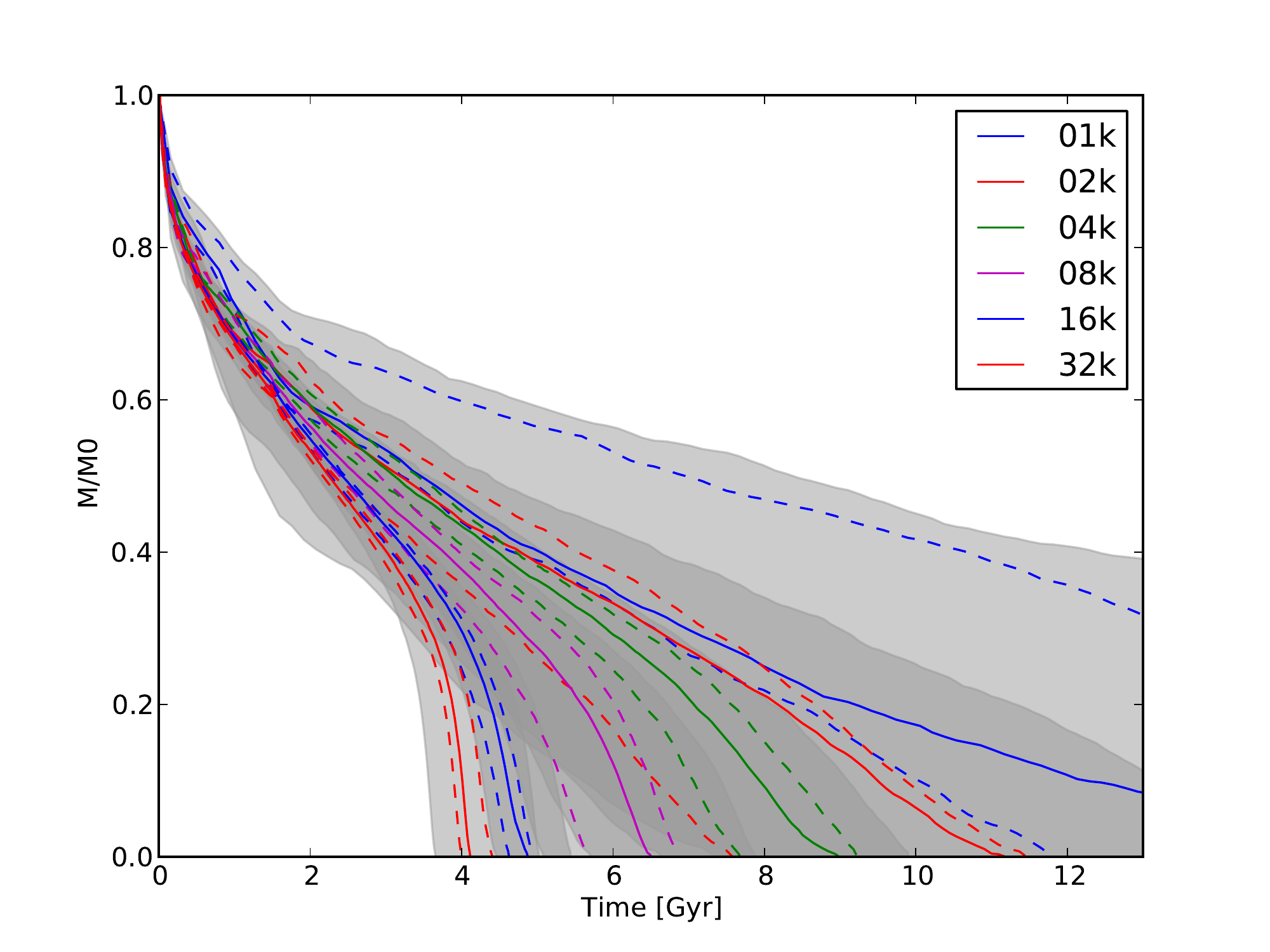}
\caption{Star cluster mass loss tracks for clusters with variable number of stars $N$.  Each $N$ has 11
random realizations.  The solid lines show the median of the runs while the dashed lines show the 25th 
and 75th percentiles.  All initial conditions are King Models with $W_0=3, \alpha=2.5, \mathrm{\rrtwo{family}}=1$.}
\label{changeN}
\end{figure*}

The intermediate region (shown as a line in Fig. \ref{theofig}) is an interesting case.  
We have seen that for ($W_0=3$, $\alpha=2.5$) the
cluster dissolves due to disruption.  However, in published plots such as \citet{F95}, there are cases within
this region of cluster dissolution due to relaxation.  We suspected that this was a change in behaviour
caused by a differing ratio of the relaxation time to the dynamical time.  In order to control this ratio,
we adjusted the number of stars $N$ for the case ($W_0=3$, $\alpha=2.5$, \rrtwo{family}=$1$).  

The result is
plotted in Fig. \ref{changeN}.  In this plot, it appears that $N=2000$ (or $N=1000$) could be either a
relaxation or disruption curve.  However, if it were to be a relaxation curve, we would expect to see a longer
lifetime for the cluster than the $N=1000$ case (comparing, for example, to Fig. \ref{21runplot}).  The $N=2000$ lifetime
is shorter than the $N=1000$ lifetime, though, which places the $N=2000$ case in the disruption region.  A similar argument
holds for the $N=1000$ case.  These small $N$ runs are, in fact, more analogous to the ($W_0=3$, $\alpha=1.5$) runs from
Fig. \ref{21runplot}.  In these cases, the cluster is dissolving in a small number of dynamical times. This set of runs 
therefore shows that all of the dissolutions for this set of parameters have a dynamical disruption character.  

One should note that dynamical disruption does not have to be a rapid process relative to a Hubble time.
While there are certainly cases of clusters being disrupted within a few tens of Myr, there are also dynamical
disruption cases where the cluster survives for several Gyr.

\subsection{Stellar Evolution Comparison} \label{secomp}

\begin{figure*}[htpb] 
\plottwo{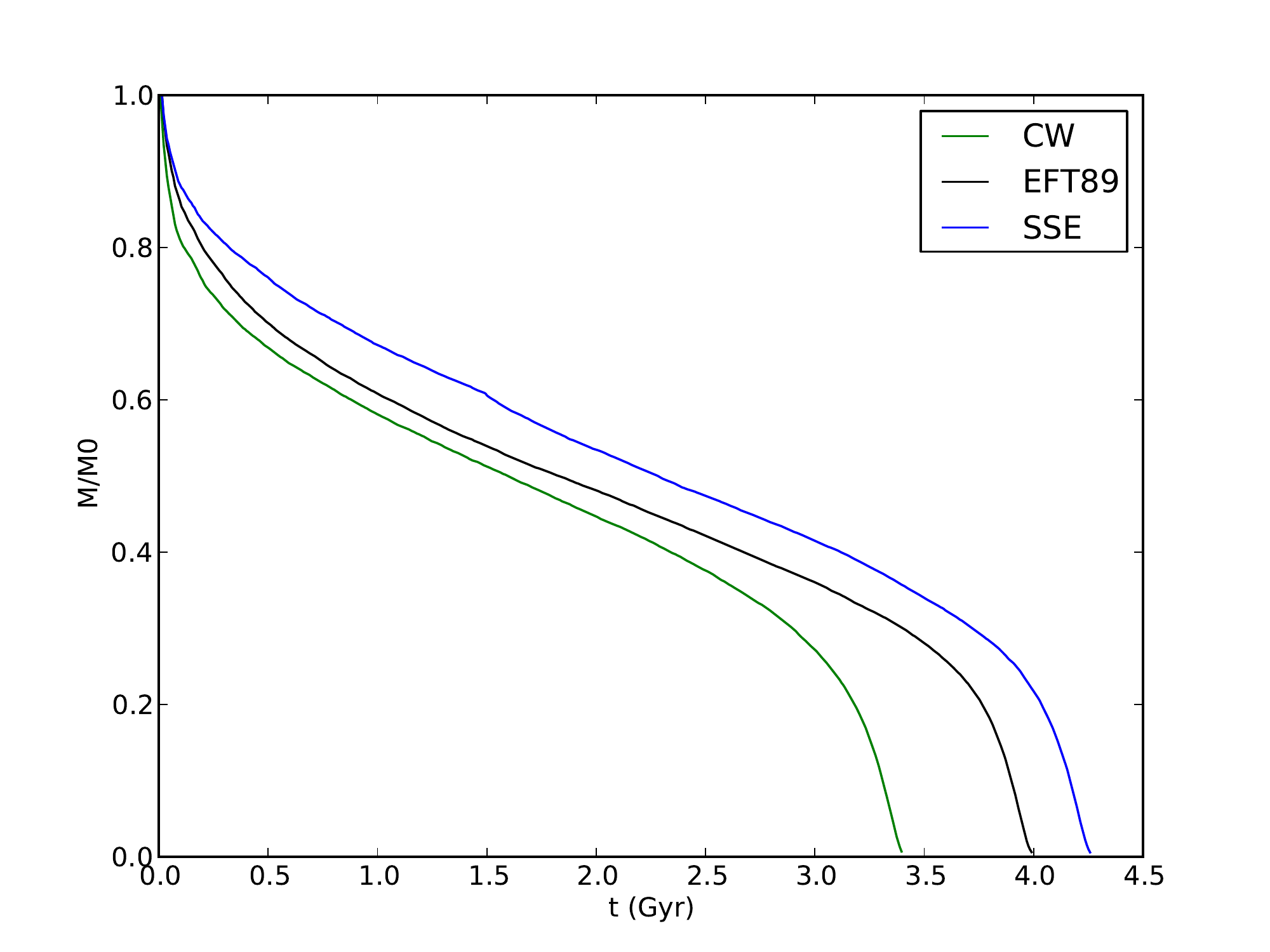}{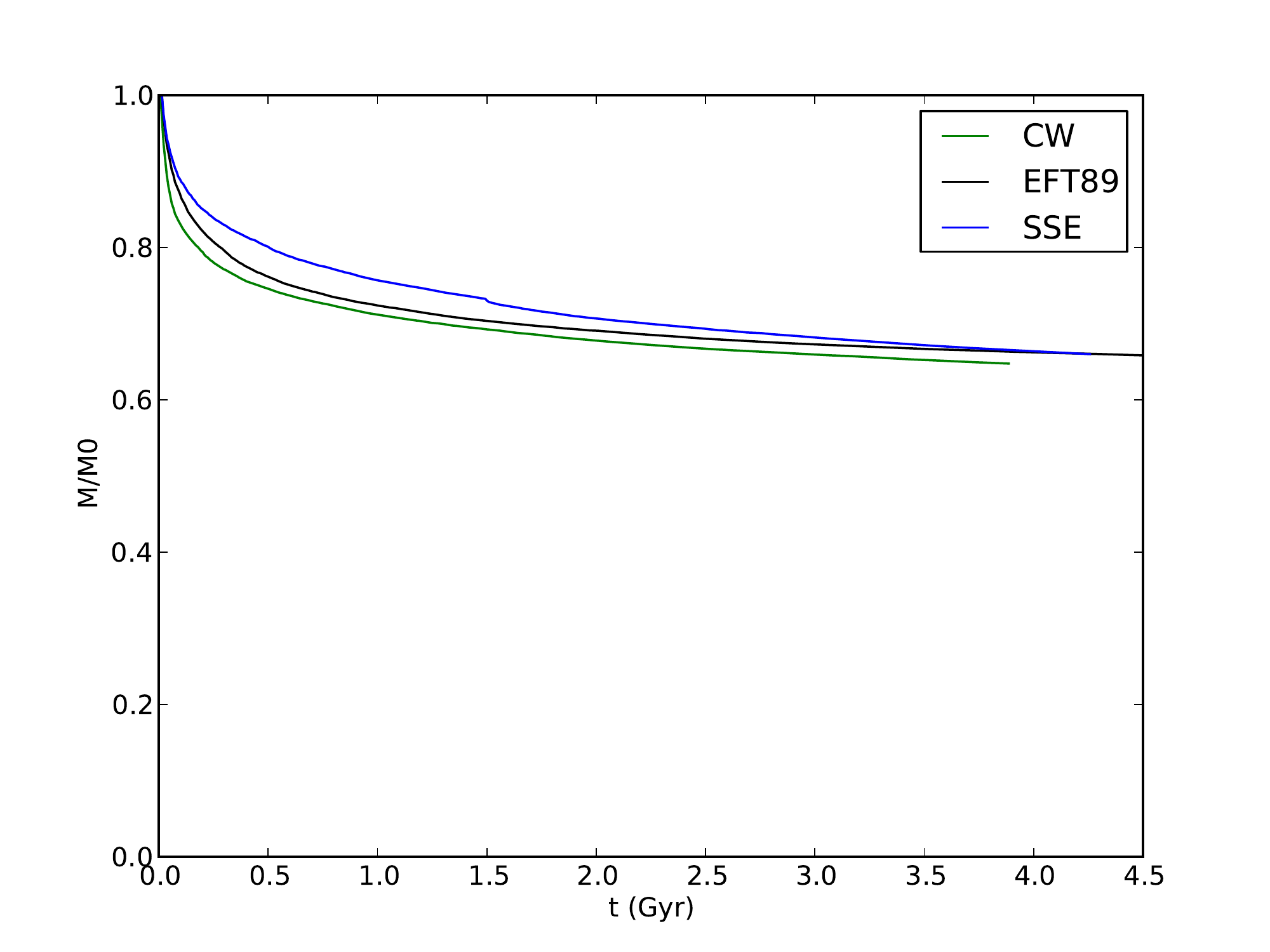}
\caption{The evolution of the mass of a cluster of N=32,000 stars,
  with initial cluster parameters $W_0 = 3$, $\alpha =-2.5$,
  $\mathrm{family} = 1$ using different stellar evolution models.  The
  left-hand frame shows the results if gravitational dynamics, stellar
  evolution and a tidal cut-off are used.  The right-hand plot shows
  the result of stellar evolution acting in isolation.  The blue,
  green, and black lines correspond, respectively, to the stellar
  evolution prescriptions of \citet{H00} (SSE), \citet{CW90} (VSSE),
  and \citet{EFT89,EFT90}.
  }
\label{compfig}
\end{figure*}

Figure \ref{compfig} shows a single $N = 32,000$ model
($W_0 = 3$, $\alpha=2.5$, $\mathrm{family}=1$) evolved using
different stellar evolution models.  {\AMUSE} easily allows switching
stellar evolution models in the same code.  We used the SSE module, as
used in the previous sections, as well as the idealized ``VSSE''
(``very simple stellar evolution'') module described previously, and
an implementation of the recipes given by \citet{EFT89}.  
All curves used were computed by {\AMUSE} using
\textsc{ph4} for gravitational dynamics.
Also included is a plot of the population
synthesis (stellar evolution only) results for these initial
conditions using the different stellar evolution models.

The small differences in the mass loss rates of the stars in the
various models are amplified by the effects of gravitational dynamics
and \rr{removal of escapers} to produce differences in overall cluster lifetime
of approximately $25\%$.  The early supernovae experienced by O 
and B stars drives the cluster toward a faster dissolution because the
relationship between the mass loss and shrinking tidal radius is self-reinforcing.
When mass is lost, the tidal radius shrinks, which may leave some stars beyond the
``edge'' of the cluster, thus leading to additional mass loss and starting the cycle over again.
These are systematic differences comparable in magnitude to
  the random run-to-run variations discussed in the previous section.

\section{Conclusions and Future Work} \label{futurework}

\subsection{Scientific Results}

In this paper, we examined whether or not \AMUSE could be compared
to published runs, in particular those of TPZ that reproduce prior work
in CW.  We also studied the source of variance between formally equivalent
simulation runs, and compared the effect of varying the stellar evolution model
on the lifetime of a star cluster.  The dissolution mechanism (disruption versus 
relaxation) for star clusters was explored.

Our {\AMUSE} runs agree with \citet{TPZ00} (which
agrees with \rrtwo{\citealt{CW90}}), when all variances are taken into account.
The sources of understood variances are: the specific differences in the random
realization of the mass spectrum, the random realization of the initial spatial
distribution of stars, and the difference in remnant masses produced by the 2000
and 2012 versions of SeBa.  These small differences are amplified by the 
dynamical interactions of stars over the cluster's lifetime.  It is not surprising
that the remnant mass prescriptions have changed over the past 12 \rrtwo{yr}, as
this has been a topic of active research in the community.

The direct comparison of stellar evolution codes has yielded
an interesting result: prescriptions which differ by no more than
a few percent in population synthesis studies can drive otherwise
identical star cluster simulations to evolve at paces that differ by
up to 25\%.  The cause is an amplification effect between mass loss
due to stellar evolution and mass loss due to \rr{stars escaping the cluster}.  As the
cluster loses mass, particularly due to early supernovae of O and B
stars, the tidal radius of the cluster shrinks and more outlying stars
are stripped \rr{from the cluster by the galaxy}.

The use of random realization to generate initial conditions was also
examined, and we found that formally equivalent (i.e.: same $W_0$, 
$\alpha$ and CW family) initial conditions do not always follow the
same evolutionary tracks.  This effect is important in those cases where the star cluster
neither dissolves rapidly (such as the loosely bound $W_0=3$ case with
a top-heavy mass function of slope $\alpha=1.5$) nor remains tightly
bound over a Hubble time (such as $W_0=7; \alpha=2.5$).  Only in the
intermediate regime (i.e.: $W_0=3; \alpha=2.5$) does this effect matter.

Variation in the evolution and dissolution time of a star
  cluster in the intermediate range is due primarily to the variation
  in masses of the most massive cluster members, but the effect of the
  differences in stellar spatial distribution due to random realization
   is also non-negligible.  As had been expected,
  the population of O and B stars dominates the early cluster evolution
  because the effects of mass loss from supernovae are amplified as the 
  cluster ages.  The specific random realization of the spatial distribution 
  also affects the overall cluster
  evolutionary track, but only at about half of the significance of the
  top-end of the mass spectrum.  This leads us to believe that the fraction
  of O and B stars alone is not sufficient to describe the variance they introduce;
  we must also consider the position within the cluster of those stars.
  
We have shown that there is a parameter-space boundary between King models
that dissolve via dynamical processes and those that dissolve via relaxation processes.
Figure \ref{theofig} shows the location of this boundary in the space of the three
relevant timescales: the stellar evolution mass loss timescale, the cluster's relaxation
timescale and the cluster's dynamical timescale.

\subsection{Computational Methods Results}

The AMUSE framework shows promise as a new method of binding together
domain-specific physics codes to form a larger simulation.  Future
work will be directed toward improving the abilities of this young
framework relative to existing monolithic codes.  Specifically, new
dynamical modules have recently been added to handle close encounters
between stars as well as the formation, evolution and destruction of
binary and multiple stars.  The addition of \rrtwo{these modules} will allow
\AMUSE to follow the evolution of a star cluster into core collapse and
beyond.

In parallel to this development, the AMUSE team has also implemented modules to provide gas dynamics (generalized hydrodynamics, in fact) and radiative transfer processes.  Using these modules, 
AMUSE should be able to perform production quality simulations including all of these components.

The modular structure of {\AMUSE} facilitates comparison of physics
modules and enables exploration of assumptions and approximations that
is difficult or impossible with other simulation codes.  We have used
\AMUSE to compare the effect of changing the stellar evolution
prescription on an otherwise identical simulation, and used that to
demonstrate that ``small'' differences between prescriptions are, in
some cases, significant in the cluster's evolution.  The ability to
directly compare individual scientific codes within a multi-physics
simulation is \rrtwo{a} novel property of a modular framework.
  
The interchangeability of modules benefits:
\begin{itemize}
  \item the users of simulation codes, who may now
  select codes from a ``menu'' of choices,
  \item those who interpret simulation results, who can
  now \rr{easily compare different implementations of the same
  underlying physics head-to-head}, and
  \item authors of new codes, who may focus their code on
  a single area of physics, knowing that it may be linked to
  a multi-physics environment easily, and that cross-validation
  of their work with existing codes can be accomplished by
  performing controlled experiments against established modules.
\end{itemize}

\acknowledgements We thank Arjen van Elteren for his help with the new
AMUSE interfaces.
This work has been
supported by NASA grants NNX07AG95G and NNX08AH15G, and by NSF grant
AST0708299.  AMUSE is developed at the Leiden Observatory, a faculty
of Leiden University, and is funded by NOVA and NWO (grants VICI
[\#639.073.803], AMUSE [\#614.061.608] and LGM [\# 643.200.503]).
Part of the work was done while the authors visited the Center for
Planetary Science (CPS) in Kobe, Japan, during a visit that was funded
by the HPCI Strategic Program of MEXT.  Another part was done while
visiting the Lorentz Center.  We are grateful to these centers for their
hospitality.

\end{document}